\documentclass[11pt,journal]{IEEEtran}
\usepackage{amsmath}
\usepackage{amssymb}
\usepackage{colortbl}
\usepackage[pdftex]{graphicx}
\usepackage{color, colortbl}
\definecolor{Row0Color}{rgb}{0.83, 0.83, 0.83}
\definecolor{Row1Color}{rgb}{0.95, 0.95, 0.96}
\definecolor{Row2Color}{rgb}{0.83, 0.83, 0.83}
\definecolor{darkgreen}{RGB}{0,100,0}
\usepackage{balance,soul}
\usepackage{color}
\usepackage[table]{xcolor}
\usepackage{siunitx}
\usepackage{url}
\usepackage{marginnote}
\usepackage{booktabs}
\usepackage{multirow}
\usepackage{cite}
\usepackage[T1]{fontenc}
\usepackage{todonotes}
\usepackage{subcaption}
\usepackage{float}
\usepackage{svg}
\usepackage{breqn}
\usepackage{circuitikz}
\usepackage{makecell}
\usepackage{pgfplots}
\usepackage{pifont}
\usepackage{cuted} 
\usepackage{bm}
\usepackage{array}
\usepackage{tikz}
\usepackage{amsmath}
\usepackage{enumitem}
\usepackage{bm}
\usepackage[normalem]{ulem}
\usepackage{booktabs, tabularx, multirow, makecell}
\usepackage{pgfplots}
\usepackage{hyperref}
\pgfplotsset{compat=1.18}
\usepgfplotslibrary{fillbetween}

\pgfplotsset{
    every axis/.append style={
        line width=0.9pt,
        tick style={line width=0.8pt},
        label style={font=\small},
        tick label style={font=\small},
        legend style={font=\small},
    }
}
\usepackage[acronym]{glossaries}
\usepackage{cuted}

\definecolor{matlow}{RGB}{192, 57, 43}
\definecolor{matmid}{RGB}{212, 136, 30}
\definecolor{mathigh}{RGB}{39, 174, 96}
\definecolor{matgray}{RGB}{180, 178, 169}

\definecolor{asc3fill}{RGB}{225, 245, 238}   
\definecolor{asc3draw}{RGB}{ 15, 110,  86}   
\definecolor{asc3text}{RGB}{  8,  80,  65}   

\definecolor{asc2fill}{RGB}{238, 237, 254}   
\definecolor{asc2draw}{RGB}{ 83,  74, 183}   
\definecolor{asc2text}{RGB}{ 60,  52, 137}   

\definecolor{asc1fill}{RGB}{241, 239, 232}   
\definecolor{asc1draw}{RGB}{ 95,  94,  90}   
\definecolor{asc1text}{RGB}{ 68,  68,  65}   

\newcommand{\gauge}[1]{%
 \begin{tikzpicture}[baseline=-0.1cm]
    \def\r{0.28}%
    \def\lw{0.8pt}%
    \ifcase#1
      \colorlet{segL}{matlow}\colorlet{segM}{matgray}\colorlet{segH}{matgray}%
      \def\needleangle{150}%
    \or
      \colorlet{segL}{matgray}\colorlet{segM}{matmid}\colorlet{segH}{matgray}%
      \def\needleangle{90}%
    \or
      \colorlet{segL}{matgray}\colorlet{segM}{matgray}\colorlet{segH}{mathigh}%
      \def\needleangle{30}%
    \fi
    \draw[segL, line width=\lw, line cap=butt]
      (180:\r) arc[start angle=180, end angle=122, radius=\r];
    \draw[segM, line width=\lw, line cap=butt]
      (118:\r) arc[start angle=118, end angle=62,  radius=\r];
    \draw[segH, line width=\lw, line cap=butt]
      (58:\r)  arc[start angle=58,  end angle=0,   radius=\r];
    \draw[matgray, line width=0.3pt] (-\r,0) -- (\r,0);
    \draw[black, line width=0.5pt, line cap=round]
      (0,0) -- (\needleangle:\r*0.78);
    \fill[black] (0,0) circle (0.03cm);
  \end{tikzpicture}%
}

\newcommand{\blocks}[1]{%
  \begin{tikzpicture}[x=1mm, y=1mm, baseline=0mm]
    \def\bw{4.0}
    \def\bh{2.5}
    \def\gap{0.8}
    \def\rad{0.4pt}
    \foreach \val [count=\i from 0] in {#1}{%
      \pgfmathsetmacro{\xpos}{\i*(\bw+\gap)}%
      \ifnum\val=1
        \filldraw[fill=black!75, draw=black!75, rounded corners=\rad]
          (\xpos,0) rectangle (\xpos+\bw,\bh);
      \else
        \filldraw[fill=white, draw=black!40, line width=0.15mm,
                  rounded corners=\rad]
          (\xpos,0) rectangle (\xpos+\bw,\bh);
      \fi
    }%
  \end{tikzpicture}%
}

\newcommand{\cmark}{\ding{51}}%
\newcommand{\xmark}{\ding{55}}%

\newcommand{\bgamma}{{\bm{\gamma}}}
\newcommand{\cnoi}{\frac{C_i}{N_0}}

\newacronym{cno}{$C/N_0$}{carrier-to-noise-density ratio}

\begin{document}

\title{GNSS Spoofing in Mobile Devices: A Survey on Impact and Countermeasures}

\author{Robert Argo,~\IEEEmembership{Member,~IEEE,} 
        Andrea Nardin,~\IEEEmembership{Member,~IEEE,} \\
        Alex Minetto,~\IEEEmembership{Member,~IEEE,} 
    and~Pau Closas~\IEEEmembership{Senior Member,~IEEE}
    \thanks{R. Argo and P. Closas are with the Dept.~of Electrical and Computer Eng., Northeastern University, Boston, MA (USA). E-mail: \{argo.r, closas\}@northeastern.edu. A. Nardin and A. Minetto are with the Dept. of Electronics and Telecommunications, Politecnico di Torino, Turin (Italy). E-mail: \{andrea.nardin, alex.minetto\}@polito.it}
    \thanks{This work has been partially supported by the National Science Foundation under Awards 1845833, 2326559 and 2530870.}
}


\maketitle

\begin{abstract}
Smartphones rely on Global Navigation Satellite System (GNSS)-based positioning for many of the functions they execute everyday. The GNSS receivers embedded in smartphones are susceptible to anthropogenic radio frequency interference attacks in the forms of jamming and spoofing due to the low-power and open-architecture signals they receive from the satellite constellations. While jamming is a practice that denies a GNSS receiver the ability to form a position, velocity, and time (PVT) solution, spoofing represents a more insidious threat by using forged satellite signals that aim at causing the victim receiver to compute a false PVT solution. The ubiquity of smartphones and the sensitive geolocation data they hold make them a primary target for malicious spoofing. However, their hardware constraints and the lack of deep visibility into the GNSS receiver processing chain create significant hurdles for effective countermeasures. Existing surveys comprehensively explore general spoofing countermeasures but fail to address these mobile-specific limitations. This article fills that gap with a novel survey focused on techniques viable within the unique constraints of smartphone architectures. Specifically, we establish a taxonomy for defining GNSS spoofing attack effects and countermeasures, provide a historical review of smartphone vulnerability characterization, and provide an overview of techniques proposed to detect and counteract smartphone spoofing threats, offering a comparative framework to weigh their respective pros and cons on mobile platforms.
\end{abstract}

%

\section{Introduction}
\label{sec:intro}
\IEEEPARstart{C}{ommercial} Global Navigation Satellite System (GNSS) receiver technology has become ubiquitous due to its low cost and effectiveness as a reliable position, velocity, and time (PVT) source. In the last decade, smartphones have benefited from this and take advantage of GNSS technology to perform many of the functions the users rely on, commonly referred to as \emph{location-based services}, including personal navigation, the communication of location during emergency calls, and the optimization of delivery logistics and ride-sharing operations \cite{zangenehnejad2021gnss}. 

\begin{figure*}[t]
    \centering
    \includegraphics[trim={0cm 1.5cm 0cm 1.5cm},clip,width=1\linewidth]{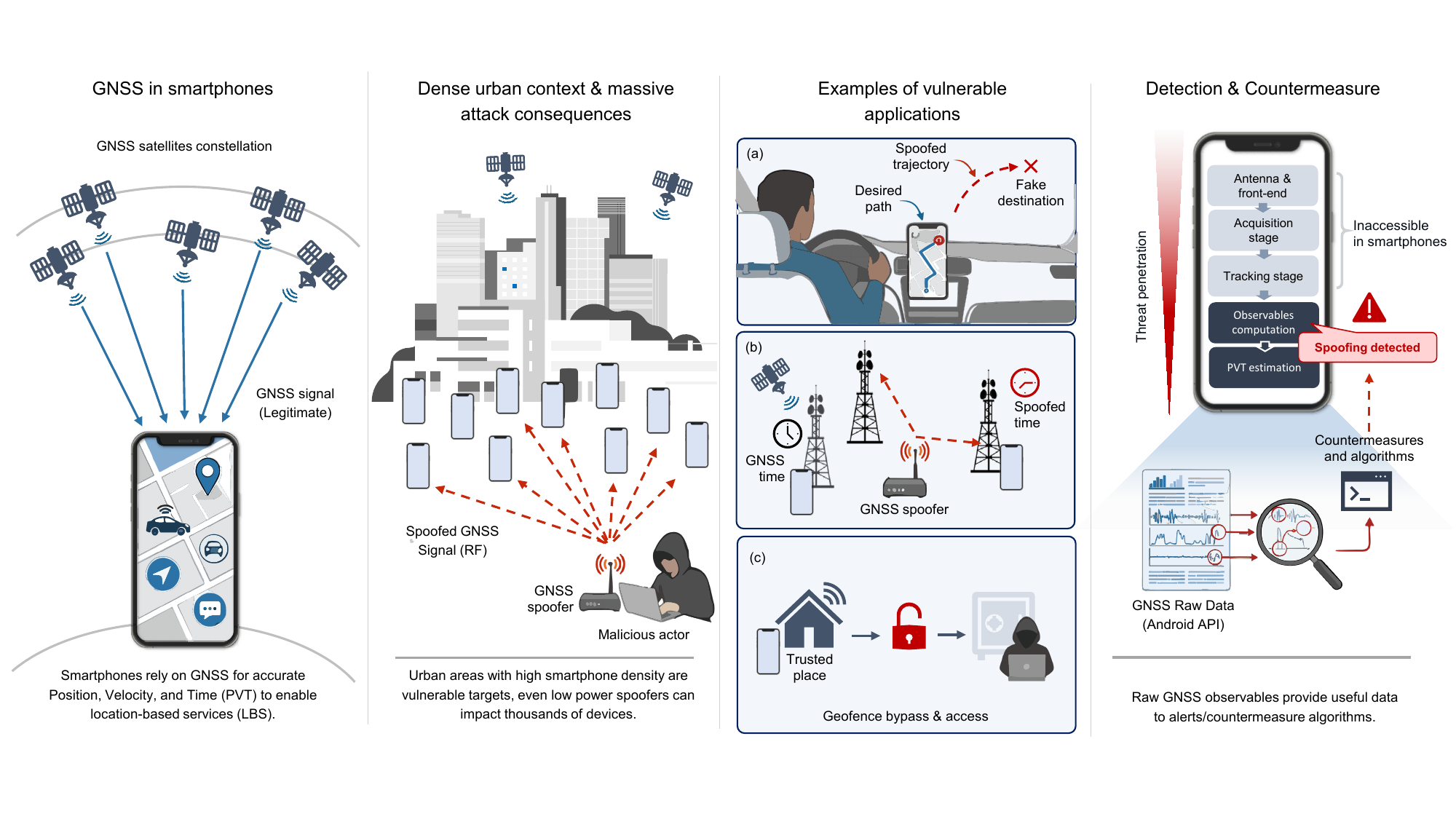}
    \caption{GNSS spoofing attacks on smartphones: a view on vulnerabilities, consequences, and detection opportunities. Subpanels illustrate representative impact domains, including (a) manipulation of position/velocity solutions in vehicular navigation systems, (b) timing perturbations affecting user devices and time-synchronized critical infrastructures, and (c) circumvention of application-level geofencing constraints through falsified location observables.}
    \label{fig:overview}
\end{figure*}

Spoofing of GNSS signals is the practice of crafting forged satellite signals that are emitted in radio frequency (RF) with the purpose of invoking false PVT estimates in a victim receiver.
Given smartphones' ubiquity and their integration of GNSS technology with personal and financial information, they offer an attractive target for a malicious actor to direct a GNSS spoofing attack against. Furthermore, in dense urban agglomerations, where smart-device densities can reach up to one million devices per square kilometer~\cite{3GPP_TR_38.913_v15.0.0}, even low-power jamming or spoofing attacks can simultaneously affect an unprecedented number of devices, as illustrated in Fig.~\ref{fig:overview}. As a result, location-based services become particularly exposed to large-scale disruptions with potentially severe financial consequences~\cite{engproc2025113042}. 

Previous works have demonstrated the effects and implications of GNSS spoofing on smartphones, including cases arising from non-malicious events, which further highlight their inherent susceptibility to such interference \cite{scott2017spoofing, zeng2017practical, rustamov2020assessment,9332076,11218853}. Among these implications, during the 2017 ION GNSS+ conference in Portland, a GNSS RF simulator that was not properly terminated resulted in numerous smartphones within about a 10 meter radius of the device lock on to the simulated GNSS signals and report the incorrect date or time~\cite{scott2017spoofing}. 
Indeed, the effect of spoofing on smartphones can also be exploited to induce a time shift, a phenomenon that has been extensively studied in the literature for critical infrastructures such as telecommunications networks~\cite{euspa_market_report_2024,euspa_timing_sync_user_requirements,xiong2022research,10381634}.
This vulnerable application is illustrated, among others, in Fig.~\ref{fig:overview}.
Researchers have also demonstrated a sophisticated spoofing attack where they were able to spoof the GNSS receiver embedded in a smartphone feeding navigation instructions to a vehicle driver~\cite{zeng2017practical}. This was done in a way that the driver was unaware he was getting false position information and navigated as much as 1\,km away from his desired destination. 
 
 Other opportunities exist to compromise personal information using GNSS spoofing of smartphones. Google's Trusted Places feature allows a user to configure their phone so that its security features are partially or fully disabled when the phone's GNSS position solution is within some geofenced area. A malicious actor with physical access to someone's phone and the ability to spoof their geofenced area could gain access to personal information in their phone without needing the correct pin or password~\cite{9332076}. Ride share drivers could spoof their phones location during a paid drive to trick their phone into detecting a longer drive than actually occurred artificially inflating fares for riders \cite{11218853}.

The susceptibility of smartphones to GNSS spoofing and its possible implications have motivated research efforts to develop countermeasure techniques that can alert users when they are potentially at risk of being in the presence of spoofing. Over the last several years this objective has benefited from the raw GNSS measurements made available through the Android API by Google in May 2016 \cite{euspa2018_gnss_raw_android}. These measurements have given researchers much lower-level access to the GNSS receiver data onboard Android smartphones~\cite{electronics8010005}, as highlighted in Figure \ref{fig:overview}. Therefore although access to intermediate observables in smartphones remains generally limited compared to standalone GNSS receivers, spoofing attacks often introduce characteristic artifacts in the raw GNSS measurements available on Android devices. These artifacts offer researchers the opportunity to develop techniques that detect and counteract spoofing when it is occurring.

\subsection{Previous Surveying Works}
\label{sec:prev_works}

In 2016, \textit{Proc. of the IEEE} published a special issue on ``Vulnerabilities, Threats, and Authentication in Satellite-Based Navigation Systems'' \cite{amin2016vulnerabilities} where renowned experts reflected about the challenges and solutions of GNSS technology at the time. More than a decade after, the field has evolved, both in terms of challenges and approaches to make GNSS more resilient. Many impactful works pointed out to the progress made in bringing high-quality GNSS chipsets to power-constrained devices like smartphones \cite{Grenier23}, as well as the relatively recent ability of users and developers to interact with the chipset \cite{fuAndroidRawGNSS2020}. That combined opened the door to a new set of vulnerabilities that need to be better understood and addressed, which is the main objective of the current article.

Existing survey papers on GNSS security and spoofing countermeasures provide valuable taxonomies of threats and detection strategies \cite{7445815,ioannides2016known,rs14194826,wang2021survey,Rados2024}, yet they typically adopt a \emph{receiver-agnostic perspective} that does not reflect the stringent constraints of mobile devices. Prior reviews frequently discuss countermeasures assuming access to raw correlation outputs, multi-band carrier phase measurements, high-rate IQ samples, disciplined oscillators, or controlled RF front-ends, none of which are available on commodity smartphones~\cite{Johansson2024}. Consequently, these studies often evaluate techniques that are \emph{inapplicable to mobile platforms}, such as sophisticated signal-quality monitoring, full multi-constellation navigation message authentication, or array-based spatial processing. Fuermore, existing surveys rarely consider the \emph{heterogeneity and fragmentation} of Android GNSS implementations—including partial or absent exposure of navigation messages, coarse clock models, and manufacturer-restricted access to chipset diagnostics~\cite{AndroidGNSSDoc}. As a result, the applicability of many methods surveyed therein to real‑world smartphone contexts remains unclear. 

Table~\ref{tab:gnss_spoofing_surveys} summarizes the aforementioned survey articles along with their main subject and specific objectives. Several of these surveys  are intentionally broad and present numerous countermeasure approaches, only some of which are, in our opinion, partially applicable to the smartphone use case~\cite{7445815}, \cite{rs14194826}, \cite{Rados2024}, \cite{Ferre2020}. Furthermore, these articles do not make clear which approaches are feasible within the unique limitations of smartphones. In one survey~\cite{Schmidt2016}, smartphones are explicitly de-emphasized with the authors' asserting that smartphones are an unlikely spoofing target due to assumed practical limitations. Subsequent studies, however, have challenged this assumption by demonstrating or assessing the feasibility and impact of spoofing attacks against smartphones~\cite{scott2017spoofing,zeng2017practical,rustamov2020assessment,9332076,11218853}. Finally, another survey specifically focuses on receiver pre-correlation data, a level of introspection unavailable in smartphones~\cite{wang2021survey}. Unlike these prior surveys, this article surveys countermeasure techniques through the lens of smartphone hardware and data constraints, offering a fresh perspective on their feasibility.



\begin{table*}[htbp]
\centering
\caption{Comparison of GNSS spoofing survey papers available in the literature.}
\begin{tabular}{l c >{\raggedright\arraybackslash}p{1.8cm} c >{\raggedright\arraybackslash}p{2cm} >{\raggedright\arraybackslash}p{2.3cm} >{\raggedright\arraybackslash}p{2.5cm}}
\toprule
\textbf{Reference (Authors)} &
\textbf{Year} &
\textbf{Main Focus} &
\textbf{Smartphone GNSS} &
\textbf{RFI-related aspects} &
\textbf{Mitigation / Detection} &
\textbf{Notes / Remarks} \\
\midrule
Psiaki \& Humphreys \cite{7445815} & 2016 & GNSS spoofing threat modeling & Not considered & Threats, attack scenarios & Receiver-level, anomaly monitoring & Broad survey focusing on COTS receivers \\
Schmidt et al. \cite{Schmidt2016} & 2016 & Threat analysis & Partly considered & Attack/defense classification & RF-/correlation-based & Smartphone case de-emphasized \\
Morales‑Ferre et al. \cite{Ferre2020} & 2019 & Intentional GNSS interference & Not considered & Jamming, spoofing & Detection, mitigation, classification & Broad interference survey for aircraft \\
Wang et al. \cite{wang2021survey} & 2021 & RF-based spoofer detection & Not considered & Jamming & RF-based methods & Focus on approaches unfeasible on smartphones \\
Meng et al. \cite{rs14194826} & 2022 & GNSS spoofing / anti-spoofing survey & Not considered & Attack modes, signal processing & Multiple strategies, ML & Broad survey; many approaches unfeasible on smartphones \\
Radoš et al. \cite{Rados2024} & 2024 & Jamming / spoofing detection survey & Not considered & Wide range of techniques & Signal processing, ML & Broad survey; many approaches unfeasible on smartphones \\
Zidan et al. \cite{article} & 2020 & GNSS Vulnerability Survey & Not considered & Intentional \& unintentional mechanisms considered & Encryption, Codeless-cross-correlation, signal statistical, antenna-based & Broad survey; many approaches unfeasible on smartphones \\
\midrule
\textbf{Argo et al., this survey} & 2026 & GNSS spoofing / anti-spoofing survey & Considered & Attack modes, signal processing & Multiple strategies &  The first smartphone-centric survey on GNSS spoofing attacks and countermeasures. Provides access to a unique smartphone spoofing dataset  \\
\bottomrule
\end{tabular}
\label{tab:gnss_spoofing_surveys}
\end{table*}

\subsection{Objectives, Scope, and Organizational Layout}

In contrast to existing surveys, the present work provides, to the best of our knowledge, the first survey explicitly constrained to countermeasures that are \emph{realistically deployable on commercial smartphones}, systematically evaluating which spoofing-detection and mitigation mechanisms remain operational given the \emph{actual GNSS observables} exposed through current GNSS APIs.  While certain unpublished works have touched upon the problem in a manner consistent with this study \cite{Johansson2024}, a formal survey adopting the proposed perspective has yet to be presented in the literature. 

In this paper, we conduct a survey of past efforts that have been made both to characterize the susceptibility of smartphones to GNSS spoofing and to explore the different techniques researchers have proposed to detect and possibly mitigate it. A heavy focus is put on Android smartphones due to the aforementioned raw GNSS measurement API available on those devices. Furthermore, we introduce a generalized taxonomy of spoofing threats and their effects, providing a unified framework that extends beyond smartphones and is applicable to GNSS receivers more broadly.

Lastly, this article features a first-of-its-kind dataset of over-the-air GNSS spoofing of a number of smartphones. To the authors knowledge, there is no other open dataset of over-the-air GNSS spoofing of smartphones, which is blocking the advancement of solutions. Section \ref{sec:dataset} provides details of this unique dataset \cite{argo2026dataset,argo2025multi}, made available for researchers to test and validate their methods.   

The rest of this paper is structured as follows: in Section \ref{sec:background} we offer a brief background on satellite signals and general GNSS receiver architectures, with particular emphasis on observable availability and receiver-chain tap points. Section \ref{sec:spoofing_effects_and_countermeasures} presents a taxonomy for categorizing the effects of spoofing attacks on GNSS receivers, introduces a framework for assessing countermeasure techniques, and provides a high-level overview of spoofing countermeasures applicable to general GNSS receiver architectures.
Section \ref{sec:techniques} reviews the literature on smartphone GNSS spoofing vulnerabilities and explores all known categories of countermeasures, offering a detailed discussion of each. We conclude the paper in Section \ref{sec:critical_comparison} with high level takeaways about the performance of different spoofing countermeasure methods, comments on possible further developments for this and related research, as well as the pointer to the available open-access dataset.

\section{Background on GNSS Receivers}
\label{sec:background}

In this section, we present a concise background on satellite navigation signals and standard GNSS receiver architectures, focusing on the observables made available at different processing stages. 

\subsection{Signal Model}
Under benign conditions, that is in the absence of spoofing signals, the received GNSS signal can be formulated as follows:      
\begin{dmath}
r(t)=\sum_{i\in\mathcal{I}^L} x_i(t;\bm{\theta}_i) + \eta(t) 
\label{eq:sig_mod}
\end{dmath}  
where $M$ is the number of in-view satellites such that 
$\mathcal{I}^L = \{1,\dots,M\}$ is the set of indices of those \textit{legitimate} satellites and $|\mathcal{I}^L|=M$, with 
\begin{dmath}
x_i(t;\bm{\theta}_i)=\sqrt{2P_i}d_i\left(t-\tau_i\right)c_i\left(t-\tau_i\right)\\
\cos{\left(2\pi\left(f_c+f_{Di}\right)t+\phi_i\right)} \label{eq:sig_mod_noiseless}
\end{dmath}  
being the received noiseless signal from the $i$-th satellite, with propagation-induced parameters $\bm{\theta}_i=(P_i,\tau_i,f_{Di},\phi_i)^\top$ denoting received signal power, code delay, carrier frequency offset, and carrier phase offset, respectively. $r(t)$ is the combined received GNSS signal, $t$ is time, $d_i\left(t-\tau_i\right)$ is the navigation message data bit stream transmitted by the $i$-th satellite, $c_i\left(t-\tau_i\right)$ is the repeated pseudorandom code sequence transmitted by satellite $i$, $f_c$ is the carrier frequency, and $\eta(t)$ is a random term characterized as additive white Gaussian noise.
    
Spoofers will emit signals with the same structure as in \eqref{eq:sig_mod_noiseless} while modifying some or all elements in its parametric representation $\bm{\theta}$ as 
well as potentially the navigation message $d_i$, for a set of pre-defined satellites. The resulting received signal would then contain a combination of both legitimate and spoofed signals, with satellite indices denoted by $\mathcal{I}^L$ and $\mathcal{I}^S$ respectively, so that
\begin{dmath}
r(t)=\underbrace{\sum_{i\in\mathcal{I}^L} x_i(t;\bm{\theta}_i)}_{\text{legitimate: } x(t;\bm{\theta})} + \underbrace{x^{S}(t;\bm{\theta}^{S})}_{\text{spoofed} } + \eta(t) 
\label{eq:sig_mod_spoofing}
\end{dmath}  
where we use the superindex $(\cdot)^{S}$ to denote that the parameters are controlled by the spoofer and we note that the spoofed signal structure is the same for both legitimate and spoofing cases except for the potential modification of the navigation bits and extra randomness added during transmission. 
Notice that we defined $x(t;\bm{\theta})$ and $x^{S}(t;\bm{\theta}^{S})$ as the received combined signals---excluding receiver noise---associated with the legitimate and spoofed transmissions, respectively, and that their parameters are a concatenation of the individual parameters: $\bm{\theta}=(\bm{\theta}_1^\top,\dots,\bm{\theta}_M^\top)^\top$ and $\bm{\theta}^S=(\bm{\theta}_1^{S,\top},\dots,\bm{\theta}_{M_s}^{S,\top})^\top$ with $|\mathcal{I}^S|=M_s$.
When sufficiently stable over time, such attacks impose drifts of the PVT solution with respect to the actual receiver location, velocity and time, as will be discussed in more detail in Section \ref{sec:spoofing_effects_and_countermeasures}.

\subsection{GNSS Receiver Basic Processing Architecture}
\label{sec:rx_basics}

Figure \ref{fig:rx_stages} shows a block diagram indicating the basic processing stages of a GNSS receiver. The combined RF satellite signal  $r(t)$ is received by an antenna which sends analog voltage values to an RF front end embedding an analog-to-digital converter (ADC). The automatic gain control (AGC) is leveraged at the analog front end prior to the ADC, where it dynamically adjusts the overall signal amplitude to match the ADC’s dynamic range. Because GNSS signals are typically well below the thermal noise floor, the AGC loop maintains an optimal ratio between the ADC’s full-scale quantization level and the noise standard deviation, thereby minimizing quantization-induced signal-to-noise degradation and daily or seasonal attenuation effects causing fluctuation in the received signal power~\cite{kaplan2017gnss}. In the presence of interference, the AGC rapidly reduces gain to prevent nonlinear effects such as ADC saturation. AGC scales are typically hardware-dependent and---particularly in Android smartphones---are ``indicative of changes in input signal power within the frequency band''~\cite{GnssAutomaticGainControl_Android}; consequently, their interpretation may be affected by opaque and device-dependent behavior~\cite{morichi2026}. Nevertheless, hardware manufacturers are making efforts to adopt more consistent reporting standards~\cite{spens2022detecting}. For the sake of clarity, this work denotes by $\alpha_{AGC}$ a consistent AGC gain applied to a conditioned intermediate-frequency signal prior to entering the ADC.

\usetikzlibrary{positioning}
\usetikzlibrary{shapes,arrows}
\usetikzlibrary{decorations.pathreplacing,decorations.markings,shapes.geometric}
\tikzset{
block/.style = {draw, fill=blue!20, rectangle, minimum height=3em, minimum width=3em},
tmp/.style  = {coordinate}, 
sum/.style= {draw, fill=white, circle, node distance=1cm},
input/.style = {coordinate},
output/.style= {coordinate},
pinstyle/.style = {pin edge={to-,t,black}
}
}

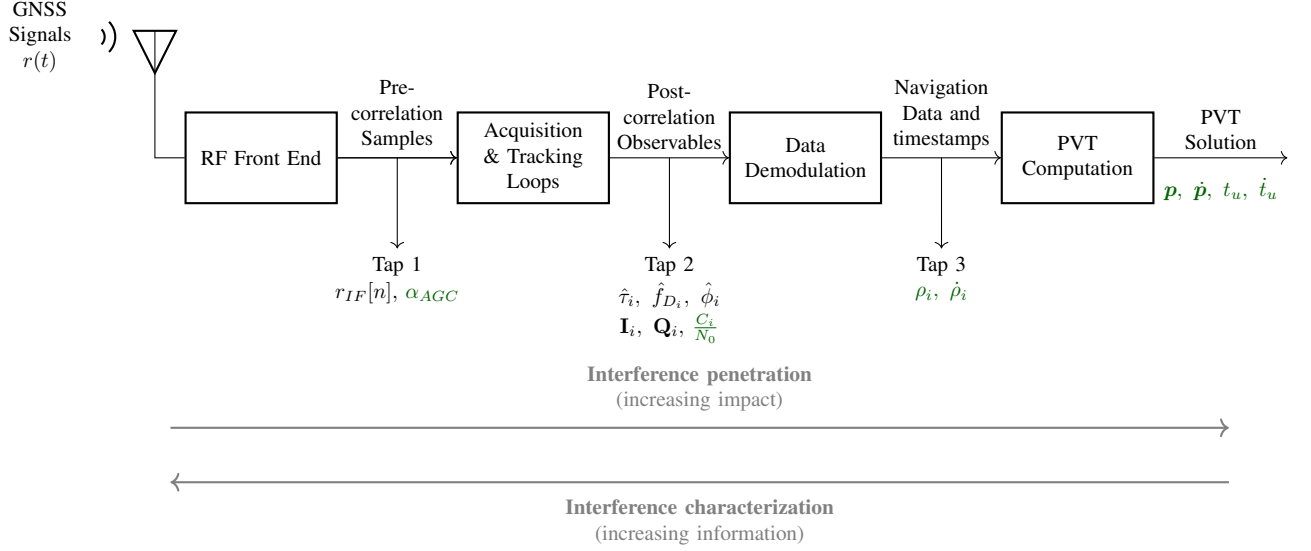
\begin{figure*}
\centering
\begin{tikzpicture}[scale=0.80, every node/.style={transform shape}, auto,font=\small, node distance=1.5cm,
        dish/.style = {circle, draw, line width=3pt, minimum size=24mm,
                     align=center},
        block/.style = {rectangle, draw, thick, minimum height=1.5cm, minimum width=2.5cm, align=center},
        output/.style = {coordinate}
                     ]

        \node [block,text width=2cm, align=center](rx){RF Front End};
        \node [block, right of=rx, text width=2cm, align=center, node distance=4.5cm](acqTrk){Acquisition \& Tracking Loops};
        \node [block, right of=acqTrk, text width=2cm, align=center, node distance=4.5cm](demod){Data \\ Demodulation};
        \node [block, right of=demod, text width=2.25cm, align=center, node distance=4.5cm](pvt){PVT \\ Computation};

        \tikzset{radiation/.style={{decorate,decoration={expanding waves,angle=90,segment length=4pt}}}}
        \draw[black, -] (rx.west)--++(0:-.5cm) node[antenna,xscale=-1] {};
        \draw[thick,radiation,decoration={angle=45}]
            ([xshift=-1.5cm,yshift=2cm]rx.west) -- +(0:.5) 
            node[left,xshift=-0.25cm,text width=2cm,align=center] {GNSS Signals 
            \\{\color{black}$r(t)$}};

        \draw [->,text width=2cm,align=center] (rx) -- node(rx2acqTrk){Pre-correlation Samples} (acqTrk);
        \draw [->,text width=2cm,align=center] (acqTrk) -- node(acqTrk2demod){Post-correlation Observables} (demod);
        \draw [->,text width=2cm,align=center] (demod) -- node(demod2pvt){Navigation Data and timestamps} (pvt);


        \coordinate (output) at ([xshift=2.2cm]pvt.east); 
        \draw [->,text width=2.5cm,align=center] (pvt) -- node{PVT \\ Solution} (output);
        \path (pvt) -- (output) coordinate[midway] (midPVT);
        \node[below=2mm of midPVT, anchor=north] {\color{darkgreen}$\bm{p},\ \dot{\bm{p}},\ t_u,\ \dot{t}_u$};

        \draw [->] (rx) -- coordinate[pos=0.5] (tap1) (acqTrk);
        \coordinate[below=of rx2acqTrk,label=below:{\shortstack{Tap 1\\{\color{black}$\displaystyle r_{IF}[n]$}$,\ ${\color{darkgreen}$\alpha_{AGC}$}  }}] (A);
        \coordinate[below=of acqTrk2demod,
label=below:{\shortstack{Tap 2\\
{\color{black}$\displaystyle 
\hat{\tau}_{i},\ \hat{f}_{D_i},\ \hat{\phi}_{i}$}\\
{\color{black}$\displaystyle 
\mathbf{I}_{i},\ \mathbf{Q}_{i}$}$
,\ ${\color{darkgreen}$\cnoi$}}}] (B);

        \coordinate[below=of demod2pvt,label=below:{\shortstack{Tap 3\\{\color{darkgreen}$\displaystyle \rho_i,\ \dot{\rho}_i $}}}] (C);
        \draw[->] (rx2acqTrk) -- (A);
        \draw[->] (acqTrk2demod) -- (B);
        \draw[->] (demod2pvt) -- (C);

        \coordinate (arrowLeft) at ([xshift=-1.5cm,yshift=-3.7cm]rx.south);
        \coordinate (arrowRight) at ([xshift=2.5cm,yshift=-3.7cm]pvt.south);

        \draw[thick,->,gray] (arrowLeft) -- (arrowRight)
            node[midway,above=4pt,align=center,text=gray]{\small \textbf{Interference penetration}\\(increasing impact)};

        \coordinate (arrowLeft2) at ([yshift=-0.9cm]arrowLeft);
        \coordinate (arrowRight2) at ([yshift=-0.9cm]arrowRight);
        \draw[thick,<-,gray] (arrowLeft2) -- (arrowRight2)
            node[midway,below=4pt,align=center,text=gray]{\small \textbf{Interference characterization}\\(increasing information)};

\end{tikzpicture}

\caption{Block diagram of baseline GNSS receiver architecture. The exposed variables identify useful observables for spoofing detection according to literature; in particular {\color{darkgreen}green} observables are currently accessible through Android smartphones APIs.}
\label{fig:rx_stages}
\end{figure*}

The ADC discretizes the analog values yielding a digital intermediate-frequency signal, as shown in~\eqref{eq:baseband},
\begin{figure*}[!ht]
\hrulefill
\normalsize
\begin{align}
\begin{split}
\label{eq:baseband}
r_{IF}[n] =
&\underbrace{\sum_{i \in \mathcal{I}^{L}} \sqrt{2P_i}\,
d_i\!\left(nT_s-\tau_i\right)c_i\!\left(nT_s-\tau_i\right)
\cos\!\left(2\pi F_{Di}nT_s+\phi_i\right)}_{\text{legitimate: }x[n]}\\
&+ 
\underbrace{\sum_{j \in \mathcal{I}^{S}} \sqrt{2P_j^{S}}\,
d_j^{S}\!\left(nT_s-\tau_j^{S}\right)c_j\!\left(nT_s-\tau_j^{S}\right)
\cos\!\left(2\pi F_{Dj}^{S}nT_s+\phi_j^{S}\right) + \eta^S[n]}_{\text{spoofed: }x^S[n]}
+\eta[n].
\end{split}
\end{align}
\hrulefill
\end{figure*}
%
%
%
where quantization, filtering, and possible multipath effects have been neglected. 
In~\eqref{eq:baseband}, $T_s$ is the sampling period, $F_{Di} \triangleq \left(f_{IF}+f_{Di}\right)$, $\eta^S[n]$ is a discrete-time random noise originally introduced by the transmission chain of the spoofer's front-end, and the notation $r_{IF}[n] \triangleq r_{IF}(n T_s),\, n \in \mathbb{Z}$ indicates a discrete-time sequence.

The signal $r_{IF}[n]$ is then sent to acquisition and tracking loops.
The acquisition and tracking loops correlate the received digital data against replicas of the satellites' pseudo-random number (PRN) codes and carrier, taking into account code delay, carrier phase, and carrier frequency offsets ~\cite{kaplan2017gnss}. 
As a result, correlator outputs are obtained in the form of (for the $k$-th correlator)
\begin{align*}
    I_{i}[k] &= \frac{1}{N} \sum_{n=1}^N r_{IF}[n]c_i(nT_s-\bar{\tau}_i[k]) \cos(2 \pi \bar{F}_{D_i} nT_s + \bar{\phi}_i) \\
    Q_{i}[k] &= \frac{1}{N} \sum_{n=1}^N r_{IF}[n]c_i(nT_s-\bar{\tau}_i[k]) \sin(2 \pi \bar{F}_{D_i} nT_s + \bar{\phi}_i)
\end{align*}

\noindent where $N$ is the number of samples considered in the correlation, $\bar{F}_{D_i}$ and $\bar{\phi}_{i}$ are, respectively, the carrier frequency and phase under test, and $\bar{\tau}_i[k]$ is the $k$-th element of a vector $\bm{\bar{\tau}}_i$ of code delays under test. A generalized vector notation of different delays is adopted here to encompass anticipated and delayed versions of the code replica---commonly referred to as the ``early'' and ``late'' replicas---as well as additional delays, such as those corresponding to the ``very-early'' and ``very-late'' in the bump-jumping method~\cite{Fine1999Trackinggnss}, which may be leveraged by different implementations. 
This array of $N_c$ delays ($|\bm{\bar{\tau}}_i|=N_c$) produces a corresponding number of correlator outputs, contained in the vectors $\mathbf{I}_{i}=\left(I_{i}[1], \dots, I_{i}[N_c] \right)^\top$ and $\mathbf{Q}_{i}=\left(I_{i}[1], \dots, I_{i}[N_c] \right)^\top$, respectively.

These values are used to estimate the code delay and carrier shift and phase of each signal according to discriminator functions and filter loops; a process that can be generalized as
\begin{equation}
\label{eq:pll_dll_estimates}
    (\hat{\tau}_{i}, \hat{f}_{D_i},\hat{\phi}_{i}) =  g(\mathbf{I}_{i}, \mathbf{Q}_{i})
\end{equation}
where $g:\mathbb{R}^{N_c}\times\mathbb{R}^{N_c}\mapsto\mathbb{R}^{3}$ is a nonlinear function encompassing the overall tracking loops operations, whose formulation typically differ according to the receiver implementation~\cite{kaplan2017understanding}. These tracking loops are typically implemented as phase/delay lock loops (PLL/DLL) \cite{Teunissen17}, generalized by Kalman filter tracking loops \cite{vila2017plls}, or computed from snapshot-based approaches \cite{fernandez2016snapshot,liu2021cloud,Grenier23}. 
Similarly, a \gls{cno} estimate is obtained as
\begin{equation}
\label{eq:cno_estimates}
    \cnoi = h(\mathbf{I}_{i}, \mathbf{Q}_{i})\ ,
\end{equation}
where $h:\mathbb{R}^{N_c}\times\mathbb{R}^{N_c}\mapsto\mathbb{R}$ can be implemented using different methods available in the literature~\cite{falletti2011cno}. Notice that the resulting power of the $i$-th satellite, $C_i$, would be an estimate of the power $P_i$~\cite{joseph2010measuring,badke2009cn0}.
Moreover, it is worth reminding that, in the presence of spoofed signals, the estimates in~\eqref{eq:pll_dll_estimates} and~\eqref{eq:cno_estimates} may reflect a mixture of spoofed and legitimate values. 

Correlator outputs are also used to demodulate the navigation message\footnote{Unlike traditional, standalone GNSS receivers, many smartphone GNSS chipsets do not handle nor provide broadcast navigation messages directly from the radio-frequency signal and instead rely on assisted data supplied through protocols such as Secure User Plane Location (SUPL)~\cite{vandiggelen2009agps}. This behavior is well documented in technical evaluations of Android smartphone GNSS performance, where it is noted that specific chipsets often do not output navigation messages at all through the Android GNSS API. As a consequence, assisted ephemeris, typically delivered via SUPL, is used to obtain satellite orbit and clock data required for positioning \cite{NavarroGallardo2021,AndroidGNSSDoc,Paziewski2021}.}. 
These estimated parameters and navigation message field values are then passed to the processing module that uses them to construct pseudorange $\rho_i$ and pseudorange rate $\dot{\rho}_i$ observables, and compute the receiver's PVT solution, which is the primary output of the receiver serving users in the end.

With this basic architecture in mind, we can consider what information is available at each stage of processing. Also shown in Figure \ref{fig:rx_stages} are three different tap points that can be considered in an open-chain architecture to access digital data in the receiver. Tap point 1 gives access to the raw digital samples that come from the {ADC}. This tap point provides all the information that exists in the received {GNSS} signals except for any information lost due to quantization and filtering. This tap point also includes any settings or \textit{pre-correlation observables} of the RF front end chip itself, such as the AGC readouts. Tap point 2 accesses the output of the acquisition and tracking loops, which include data such as the correlators' outputs and code delay and carrier frequency offset estimates. 
At this stage of processing, the available information is reduced. The receiver provides parameters from the matched filter output of the received signal against a subset of the PRN codes it is tracking. The exact information depends on the receiver implementation, but some reduction is inevitable due to processing constraints.

Tap point 3 provides access to the pseudorange and pseudorange rate observables as well as the demodulated navigation message field values\footnote{
 Qualcomm Snapdragon devices historically expose GPS L1 C/A navigation messages on selected models (e.g., Snapdragon~845/855, 888, 8~Gen~1), though support is inconsistent across manufacturers and firmware builds; in many cases, only GPS is available, with Galileo, BeiDou, and QZSS navigation messages remaining inaccessible \cite{NavarroGallardo2021,GoogleGNSSAPI}. Pixel devices using Google Tensor (G1--G3) provide partial GPS navigation message output depending on firmware revisions \cite{GooglePixelMeasurements}.

Broadcom BCM47755/BCM47765, the dual-frequency GNSS chipsets used in early mass-market dual-frequency smartphones, are known to support navigation message extraction, and early engineering devices (e.g., Xiaomi~Mi~8) briefly exposed these messages; however, this functionality is often disabled by OEMs in production firmware \cite{Wen2020,BCM47755Datasheet}.

Mediatek (Helio/Dimensity) chipsets implement onboard navigation message demodulation internally, but Android-level access to \texttt{GnssNavigationMessage} is almost always disabled in commercial devices, consistent with independent evaluations of raw GNSS measurement availability \cite{Paziewski2021,AndroidGNSSDoc}}.
\noindent The last opportunity for receiver introspection is in the receiver PVT output itself, which may offer only limited information of the original received signal.
In general, a smartphone developer has access to the PVT output exposed at the application layer. Thanks to the raw GNSS measurements API on Android devices~\cite{AndroidRawGNSS}, the developer may also access a variety of measurement fields which can be essentially reduced to a few observables---such as the AGC level, the $C/N_0$, the pseudorange rate, and from the timing information, the pseudorange itself~\cite{euspa2018_gnss_raw_android}. 
At the time of writing, to the best of our knowledge, no commercially available smartphone chipset (nor publicly exposed API) allows access to raw digital signal samples, correlator outputs, or other lower‐level intermediate observables found in dedicated GNSS receivers. The accessible observables in an Android smartphone are highlighted in green in Figure~\ref{fig:rx_stages}. Importantly, the subset of observables available in smartphones precludes researchers from exploring known spoofing countermeasure techniques that utilize other measurements included in black text in Figure~\ref{fig:rx_stages}. Arrows indicating increasing interference penetration and interference characterization are indicated below the block diagram in Figure~\ref{fig:rx_stages}. \textit{Interference penetration} refers to the extent to which interference energy permeates the receiver chain, where later stages of the receiver chain correspond to deeper penetration and  higher impact to the user space. In contrast, \textit{interference characterization} refers to the level of information available to assess receiver state, where earlier stages in the receiver chain correspond with a higher level of information.

\subsection{Position, Velocity, and Time calculation}
\label{sec:pvt}

The ultimate objective of a GNSS receiver is to compute its PVT, a real-valued \textit{state vector} $\bgamma$ with an associated uncertainty typically expressed through its error covariance. This may involve calculation of the three-dimensional user position $\bm{p}=(x_u, y_u, z_u)^\top$; its velocity $\dot{\bm{p}}=(\dot{x}_u,\dot{y}_u, \dot{z}_u)^\top$; and the receiver clock offset and clock drift $t_u$ and $\dot{t}_u$, respectively. In common applications, position and clock offset are reported, with other quantities like acceleration sometimes estimated as well.

This solution is typically computed once the set of observables from a minimum of $M\geq4$ satellites $\mathcal{O}\triangleq \left\{ \rho_i,\ \dot{\rho}_i \right\}_{i=1}^M$ are obtained, as well as relevant ephemeris data such as satellite orbital parameters and transmission times \cite{kaplan2017understanding,morton2020position}. Other direct-positioning approaches are possible, which do not require intermediate per-satellite processing thus enhancing the receiver sensitivity to operate in challenging environments \cite{closas2017direct}. However, this is left outside of this survey in order to focus on what is currently available on smartphones.

The details of the PVT computation can be found in practically all textbooks in the GNSS literature. Essentially, a geometrical problem is solved given the computed observables from the tracked satellites. This results in a (mildly) nonlinear problem that can be solved through a number of estimation alternatives. Most popular solutions are $i)$ least squares (LS) for single point positioning (SPP) solutions, where a set of observables is used to provide an estimate of the receiver's PVT at a single time point; and $ii)$ Bayesian estimators such as Kalman filters for continuous PVT tracking assuming a dynamical model for the receiver's PVT and proper prior information about the receiver state. The latter is also popular due to its practical use in sensor fusion setups, for instance with inertial measurement units (IMUs) \cite{dardari2015indoor}. Other alternatives are possible, including recent approaches where artificial intelligence is leveraged to account for additional sources of information \cite{Imbiriba24apbmtaes,Shlezinger25aikf}, although LS and KF are arguably the most common approaches to date. 

A PVT solver can be generally interpreted as a function of the available observables $\mathcal{O}_k$ at a given epoch $k$ which produces the desired PVT $\bgamma_k = (\bm{p},\ \dot{\bm{p}},\ t_u,\ \dot{t}_u)^\top$ estimate $\hat{\bgamma}_k$. Additionally, the estimator requires a set of valid ephemerides $\mathcal{E}$ such that satellite positions and velocity may be computed. That is,
\begin{align}
   \hat{\bgamma}_k = \Xi \left( \mathcal{O}_k, \mathcal{E}, \hat{\bgamma}_{k-1}, \mathcal{S}_k \right) \label{state_estimate}
\end{align}
where prior PVT estimates $\hat{\bgamma}_{k-1}$ may be used in the process (e.g., when employing filtering solutions), as well as information from other sensors $\mathcal{S}_k$ \cite{dardari2015indoor}.

\section{GNSS Spoofing Effects Taxonomy and General Countermeasures}
\label{sec:spoofing_effects_and_countermeasures}

\subsection{Receiver Effects of Spoofing}

In this section we briefly explore the effects of spoofing on a receiver and present a framework for assessing spoofing effectiveness. The framework we developed offers a novel way of categorizing spoofing by adopting a \textit{receiver effects-focused approach}, categorizing attacks by their level of impact on the receiver processing chain. Other established works adopt a classification framework grounded in the temporal and spectral alignment between counterfeit and authentic signals. This approach is widely recognized as a reference standard in both academic research and industrial technical analyses \cite{humphreys2008assessing,wu2025detection}. However, it focuses on spoofing at the signal level, without consideration for the complexities often found in the victim receiver chain. Furthermore, recent socio-technical frameworks have extended traditional classifications by incorporating the \textit{Bell–Whaley Theory of Deception}, distinguishing attacks according to strategies of dissimulation and simulation \cite{bell2017cheating}. Within this evolving landscape, artificial intelligence is increasingly considered not only as a defensive mechanism but also as a potential enabler of adaptive and intelligent attack generation. This development further reinforces a need for updated classification criteria capable of addressing GNSS security challenges in highly automated and AI-driven operational contexts \cite{marata2025trends}.

Spoofing attacks aim to induce fake observable construction upon illegitimate signal reception. The realized effect on the victim receiver processing chain for a particular scenario will depend on multiple factors, such as spoofer sophistication, spoof signal power incident on the victim receive antenna relative to authentic signal power, and victim receiver resistance to spoofing, to name a few. In regards to spoofer sophistication, attacks are commonly categorized as targeted or non-targeted. Targeted attacks are specifically engineered to impose a predetermined false trajectory or timing bias on the victim receiver, whereas non-targeted attacks aim at broader service disruption, such as denial of service or the induction of random position deviations \cite{11048492}. These compounded factors make prescribing a specific outcome to a particular spoofing scenario highly complex and often untenable. To accommodate this complexity, we generated a framework for  categorizing attacks that broadly encompasses all these nuances using a receiver effects-oriented approach.

In the following, we introduce a taxonomy to classify spoofing effects according to the composition and consistency of the observable set $\mathcal{O} = \mathcal{O}^{L} \cup \mathcal{O}^{S}$, where the subscript $k$ is omitted for readability, and $\mathcal{O}^{L}$ and $\mathcal{O}^{S}$ denote the subsets of legitimate and spoofed observables, respectively. Notice that a receiver can have a mix of both, depending on which satellites have been successfully spoofed. This taxonomy enables the identification of distinct receiver-level scenarios, which in turn facilitate the assessment of the attack’s effectiveness.

Notice that, by fabricating the data component in generated spoofed signals, a spoofer may induce, alongside fake observables, also a set of fake ephemerides ($\mathcal{E}^{S}$) to be used within the receiver PVT routine. For the sake of simplicity, the rest of the paper focuses only on fabricated observables, without loss of generality.

\begin{figure*}[t]
    \centering
    \includegraphics[trim={5cm 0 5cm 0},clip,width=0.9\linewidth]{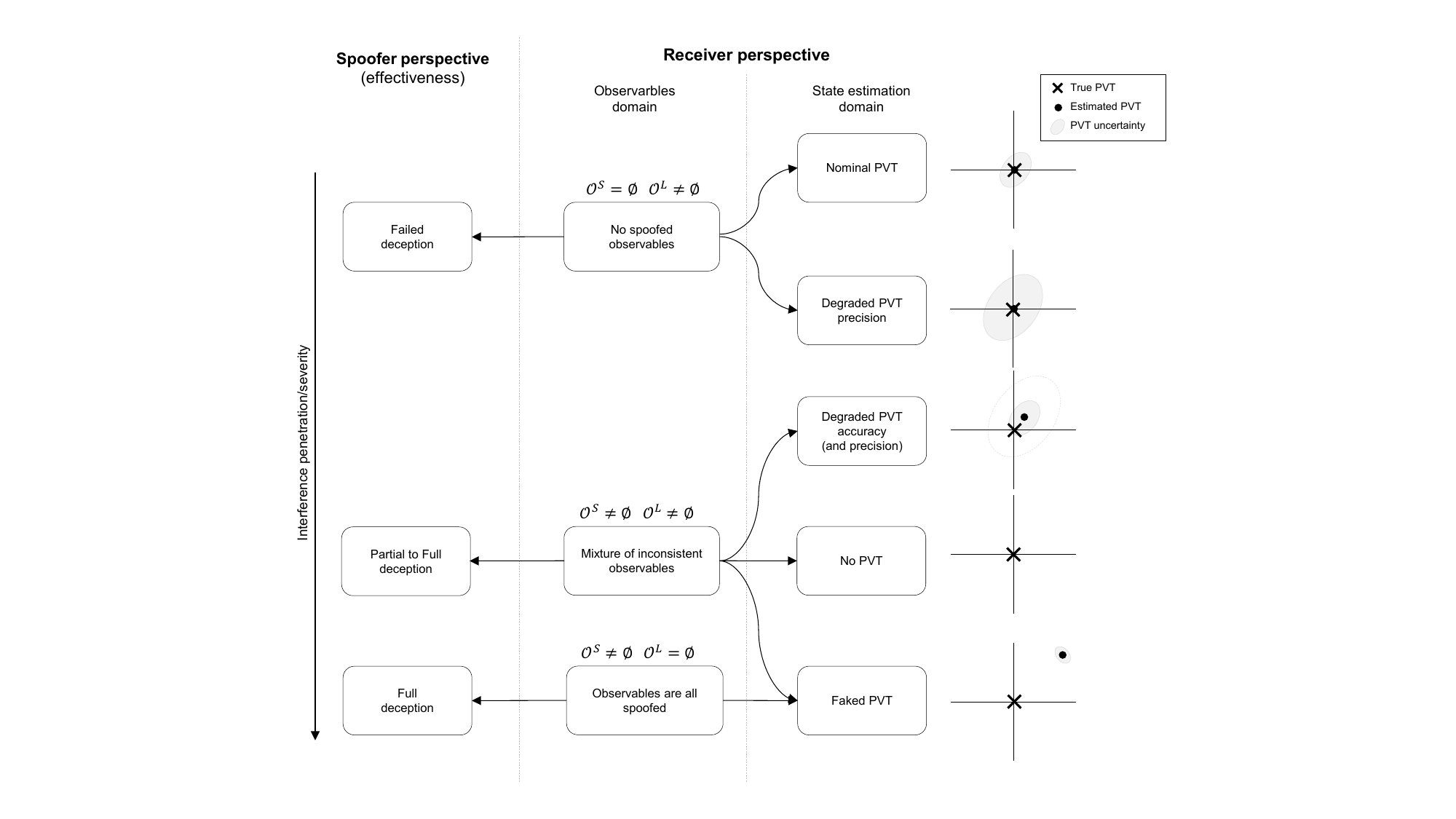}
    \caption{Block diagram presenting the impact on observables and receiver state estimation of different spoofing scenarios, according to the composition and coherence of observable sets.}
    \label{fig:rx_effects}
\end{figure*}
Figure \ref{fig:rx_effects} illustrates the developed taxonomy. We present spoofing effectiveness from two different perspectives, the attacker perspective and the receiver perspective. Within the receiver perspective, we distinguish the observable domain and the state estimation (PVT) domain. When the effectiveness of spoofing is low, $\mathcal{O}^{S} = \emptyset$ and therefore $\mathcal{O}=\mathcal{O}^{L}$. Then, if $\mathcal{O}^{L} \neq \emptyset$, the receiver only processes legitimate measurements yielding a nominal PVT solution. As spoofing power becomes non-negligible but remains insufficient for a coherent takeover (i.e., to force the receiver to track spoofed signals), the measurement model is affected through an increase in the noise term, 
resulting in degraded estimation precision, analogous to jamming. Although the performance may be degraded, both cases can be considered as a failed deception from the attacker's perspective.

Partial to full deception arises when legitimate and spoofed observables coexist, i.e., when both $\mathcal{O}^{S}$ and $\mathcal{O}^{L}$ are non-empty sets, leading to a mixture of generally inconsistent measurement set. In this regime, the estimator may converge to a biased state with increased covariance, fail to converge to a consistent solution, or even lead to a fully forged PVT solution. The exact behavior of the receiver and PVT solution in this case is in general difficult to predict and dependent on the conditions of the exact scenario.

Full deception corresponds to the limiting case $\mathcal{O} = \mathcal{O}^{S}$, in which the receiver does not gather legitimate observables and the PVT estimator converges to a controlled but internally consistent false state. 

The overall resulting PVT outcomes—ranging from nominal operation to degraded precision, biased estimates, loss of solution, or fully falsified PVT—are depicted in Figure~\ref{fig:rx_effects} in the state-estimation domain.

\subsection{Countermeasure Technique Assessment Framework}
 
\begin{figure}[t]
  \centering

  \resizebox{\columnwidth}{!}{%
  \begin{tikzpicture}

    \filldraw[fill=asc3fill, draw=asc3draw, line width=0.5pt]
      (0,0) ellipse (6.5cm and 3cm);
    \node[font=\small\bfseries, text=asc3text] at (0, 2.5cm) {Mitigation (ASC3)};

    \filldraw[fill=asc2fill, draw=asc2draw, line width=0.5pt]
      (0,-0.15cm) ellipse (4.4cm and 2cm);
    \node[font=\small\bfseries, text=asc2text] at (0, 1.35cm) {Identification (ASC2)};

    \filldraw[fill=asc1fill, draw=asc1draw, line width=0.5pt]
      (0,-0.35cm) ellipse (2.2cm and 1.1cm);
    \node[font=\small\bfseries, text=asc1text] at (0, -0.35cm) {Detection (ASC1)};

  \end{tikzpicture}
  }
  \caption{Relationship between anti-spoofing categories (ASCs) and their nested effectiveness measures.}
  \label{fig:asc-diagram}
\end{figure}
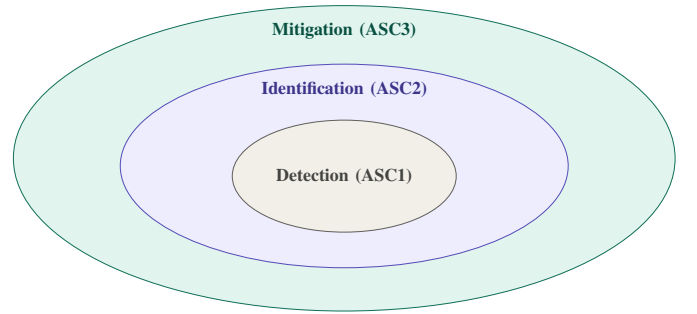

All countermeasure techniques have the same essential goal: minimizing the cardinality of $\mathcal{O}^{S}$. The techniques differ in how effective they are in achieving this goal. We define 3 Anti-Spoofing technique Categories (ASCs) based on their effectiveness: 
\begin{itemize}
    \item[$\cdot$] \textbf{ASC1}. Detection of the conditions $\mathcal{O}^{S} \neq \emptyset$ \label{cat1}
    \item[$\cdot$] \textbf{ASC2}. Identification of the specific elements in $\mathcal{O}^{S}$ indicating which satellites are spoofed \label{cat2}
    \item[$\cdot$] \textbf{ASC3}. Mitigation and restoration of the condition $\mathcal{O} = \mathcal{O}^L$ (which implies $\mathcal{O}^{S}=\emptyset$) that the receiver is not deceived with spoofed satellites \label{cat3}
\end{itemize}
The categories are a nested hierarchy with effectiveness increasing from {ASC1} to {ASC3}, as depicted in Fig. \ref{fig:asc-diagram}. Techniques in higher categories can implicitly perform an equivalent function to techniques in lower categories.

The consequence of techniques on the PVT solution can be understood by referring back to Figure \ref{fig:rx_effects}. Employing any countermeasure technique will generally enable moving upward in Figure \ref{fig:rx_effects}. For example, a receiver in the ``Faked PVT'' state that employs a countermeasure technique of category \textbf{ASC1} can detect it is being spoofed ($\mathcal{O}^{S} \neq \emptyset$). This information can then be used to reduce the amount of false information in the produced PVT solution (for instance, ceasing to produce a PVT solution and moving upward to the ``No PVT'' state). We refrain from prescribing specific PVT outcomes for a given technique because the PVT state can be affected by factors outside of the technique itself (for example, receiver complexity). For this reason we categorize each technique presented in this paper, and leave as an exercise for the reader the result of the PVT solution for a specific scenario of interest.

\subsection{General Countermeasure Approaches for GNSS Spoofing}

In this section we offer a brief review of countermeasure approaches available in literature for the general GNSS receiver (not exclusively for smartphones). As a part of this review we assess each approach using the proposed framework in this section. Countermeasure techniques for GNSS spoofing signals have been extensively studied in previous literature \cite{borio2016impact,8882350,7445815,junzhi2019research,schmidt2018software}. These surveys explored spoofing countermeasure approaches that made varying assumptions about which data tap points can be leveraged. As we will see, the level of information available from the receiver processing chain dictates what approaches are possible.

Smartphones present a unique challenge because, historically, they have offered limited data from the GNSS chip, in the most limiting cases only providing the time series of the receiver's position estimates \cite{zangenehnejad2021gnss}. Furthermore, smartphones have restrictions on the hardware they have available due to cost constraints and also must support a wide variety of use cases in their role as general computing devices \cite{s22249879}. Due to these restrictions, many of these studies have limited applicability to the smartphone use case. Some of the assumptions made that discount techniques from being used in the smartphone case include requiring hardware not available on most smartphones, requiring the receiver is stationary, requiring the receiver follow a predetermined movement pattern, requiring processing of digitally signed signals, or access to the data tap points in the receiver processing chain unavailable in many smartphones. 
Nevertheless, there are many techniques that operate on the data available in smartphones and have been used to develop spoofing countermeasure algorithms within the confines of what smartphones offer.
Table \ref{tab:methods} summarizes common spoofing countermeasure methods employed along with their limitations.

\begin{table*}[t]
\caption{Summary of general spoofing countermeasure method families in GNSS receivers. Applicability to smartphones and limitations of techniques are noted.}
\label{tab:methods}
\centering
\renewcommand{\arraystretch}{1.3}
\setlength{\tabcolsep}{4pt}

\begin{tabularx}{\textwidth}
{c p{2.9cm} c c p{5.1cm} c}
\toprule
\textbf{Domain} &
\textbf{Countermeasure method} &
\textbf{\makecell{Available in\\ smartphones}} &
\textbf{\makecell{Countermeasure \\category}} &
\textbf{Comment} &
\textbf{References} \\
\midrule

\multirow{3}{*}{Hardware-Based}
& Antenna array
& \xmark
& \blocks{1,1,1}
& GNSS antenna arrays not available on smartphones
& \cite{7444116,7442765,7444141,7444122,7440776} \\

& Correlation with auxiliary sensor data
& \cmark
& \blocks{1,1,0} 
& Inertial sensor data widely available on smartphones
& \cite{lee2019analysis,7441992} \\

& AGC monitoring
& \cmark
& \blocks{1,0,0} 
& Data readily available in Android API
& \\

\midrule
Cryptographic
& Signal encryption
& \xmark
& \blocks{1,1,1} 
& Processing of encrypted signals not currently available on smartphones
& \cite{7445815,odriscoll2025assisted,7442769} \\

\midrule
\multirow{3}{*}{Pre-correlation}
& MLE-based distortion
& \xmark
& \blocks{1,1,1} 
& Requires access to pre-correlation data unavailable in smartphones
& \cite{wang2017gnss} \\

& Multiple correlators per channel
& \xmark
& \blocks{1,1,1} 
& Requires extensive architectural changes to current low-level GNSS receiver algorithms
& \cite{nielsen2011gnss} \\

& Interference cancellation
& \xmark
& \blocks{1,1,1} 
& Requires access to pre-correlation data unavailable in smartphones
& \cite{sathaye2020semperfi} \\

\midrule
\multirow{8}{*}{Post-correlation}
& \gls{cno} monitoring
& \cmark
& \blocks{1,1,0} 
& Data readily available in Android API
& \makecell{\cite{minetto2023detecting}, \cite{rustamov2023improving},\\
\cite{borio2016impact}, \cite{valles2015interference}} \\

& Signal quality monitoring
& \xmark
& \blocks{1,1,0} 
& Signal quality metric not available in Android API
& \cite{manfredini2015signal,cavalieri2010,Pini2011SQMSpf} \\

& RAIM
& \xmark
& \blocks{1,1,0} 
& Not currently available in smartphones
& \cite{ledvina2010line} \\

& Multi-observable clustering
& \cmark
& \blocks{1,1,0} 
& Performance dependent on training data and raw data available from GPS chip
& \cite{semanjski2020use,chen2022gnss,gutierrez2025mapt} \\

& Doppler shift monitoring
& \xmark
& \blocks{1,1,0} 
& Possible in principle, smartphone use case not amenable to this approach
& \cite{calvo2020short} \\

& Pseudorange-Doppler consistency
& \cmark
& \blocks{1,1,0} 
& Requires observable inconsistency between pseudorange and Doppler measurements
& \cite{tao2019gnss} \\

& Correlation with other GNSS signals
& \cmark
& \blocks{1,1,0} 
& Requires receiver process signal not transmitted by spoofer (e.g.\ L5 frequency)
& \cite{7445815} \\

& Clock offset/drift monitoring
& \cmark
& \blocks{1,1,0} 
& Dependent of oscillator class spoofer sophistication
& \makecell{\cite{jafarnia2013pvt}, \cite{10802957},\\ \cite{chen2020moving}, \cite{shang2020gnss},\\ \cite{truong2023characterization},  \cite{zhou2024clock}} \\

\bottomrule

\multicolumn{6}{@{}p{3in}}{\textbf{Countermeasure category key}}\\
\multicolumn{6}{@{}p{7in}}{\blocks{1,0,0} ASC1 \qquad \blocks{1,1,0} ASC2 \qquad \blocks{1,1,1} ASC3 }
\end{tabularx}
\end{table*}

\section{Smartphone-specific Spoofing Impact and Countermeasures}
\label{sec:techniques}
In this section, we discuss spoofing impact on smartphones and the different major approaches that researchers have proposed to counteract spoofing. As anticipated, spoofing detection is a prerequisite for mitigation, particularly when it allows identification of which signals are being spoofed. Moreover, effective detection provides a valuable trigger in multi-source, sensor-fused navigation systems, enabling the receiver to discard compromised GNSS data and maintain robust positioning in the presence of attacks. In the ensuing subsections, we first explore the susceptibility of smartphones to GNSS spoofing and subsequently investigate the primary classes of spoof countermeasure algorithms applied to smartphones: AGC \& \gls{cno}-based methods, temporal correlation-based methods, inertial-based methods, crowdsource-based methods, multi-data source-based methods, and multi-frequency-based methods. In our investigation of each method type, we comment on the effectiveness of each using our aforementioned assessment framework.

\subsection{Spoofing Impact on Smartphones}
\label{sec:impact}

Numerous independent research groups have conducted RF testing to assess smartphone resiliency to spoofing \cite{argo2025multi}. In general, the observed effects and the associated vulnerability are primarily determined by the particular combination of front-end assembly and chipset characterizing each smartphone architecture, as well as by the experimental conditions. The researchers' experiment setups are typically not identical, but the same high-level conclusion is often drawn: smartphones are susceptible to GNSS spoofing and exhibit limited resistance to abruptly changing their PVT solutions away from historic or complementary data. Early research that precedes the release of the raw GNSS measurement access in Android showed smartphone susceptibility to GNSS spoofing~\cite{huang2015low},\cite{valles2015interference}. More recent research using the GNSS raw measurement API in Android devices corroborates this early research. In \cite{ceccato2018exploiting} the authors look at an extensive set of devices over various spoofing approaches. They demonstrate a wide susceptibility to spoofing in both the ranging codes and the navigation message on most of the smartphones tested. In \cite{rustamov2020assessment}, the authors assessed the vulnerability to spoofing attacks of three different Android smartphones with a spoofer operating in the GPS L1 band. The authors initially found the phones had limited susceptibility to spoofing attacks. In their experiments they observed the smartphones would track their forged signals, but would often not include those tracked signals in their GNSS solution. The high level behavior observed was either a denial of GNSS-based geolocating or no effect on the phone's reported position and time. The authors hypothesized that this robustness came from a combination of authentic L5 signals that the phones were able to continue tracking, as well as network-based positioning that was still available from the phones' cellular and/or Wi-Fi capabilities. 
Subsequent research performed by the same  team expanded on this work and showed a larger vulnerability to GNSS spoofing in smartphones than originally indicated in their first study \cite{rustamov2023improving}. This study investigated a much wider set of smartphone devices testing 18 different devices instead of the first study's original 3 devices. Furthermore, the spoofing scenario was adjusted to investigate the smartphones' behavior to spoofing in a more trying environment. To this end, the authors reduced the position difference between the true and spoofed position from 144\,km to 12\,km. They also included a preemptive jamming attack to cause loss of lock of legitimate {GNSS} signals. Under this different set of experimental conditions and considering a wider set of devices, the authors found a much higher susceptibility of the phones to reporting the spoofed location. Other research groups have independently demonstrated smartphones' susceptibility to GNSS spoofing. Researchers from the University of Colorado Boulder demonstrated experiments showing smartphone positions spoofed via their GNSS chipsets~\cite{miralles2020robust}. Recent research to the writing of this paper has confirmed smartphones' continued susceptibility to spoofing. In \cite{mcburney2024empirical} the authors reported data of a real life spoofing event in Haifa, Israel where several smartphones were spoofed into reporting incorrect position data.
\linebreak
While there are many studies demonstrating the susceptibility of smartphones to spoofing as well as proposed methods to detect spoofing, there is little evidence of any historic or emerging capability in smartphones to resist or detect spoofing. At the time of writing, modern Android devices do not natively support GNSS spoofing countermeasures. While the operating system provides standard location services, it lacks integrated mechanisms to verify signal authenticity. However, starting from Android 7, the Raw GNSS Measurements API allows applications to access pseudorange, Doppler, and signal-to-noise data for visible satellites. This enables the development of software-based spoofing detection strategies, often combining GNSS data with inertial sensors or network-based location information to identify anomalous or inconsistent signals indicative of spoofing attacks.

\subsection{AGC \& ${C}/{N_0}$ Detection Algorithms}
\label{sec:agc_cn0_sec}

Perhaps the most common approach to detecting spoofing in smartphones uses the measurements obtained from AGC and \gls{cno} \cite{valles2015interference}. An implementation of this spoofing detection method is included in the current Android GNSSLogger app, though at the time of writing this functionality is experimental and not yet extended to the operating system by Google. 

This type of approach takes advantage of the fact that a spoofing signal will typically increase the RF energy impinging on the GNSS antenna aperture of the smartphone by an amount that produces observable effects in the AGC and/or reported \gls{cno}. 
This decrease in AGC gain in tandem with an increase in \gls{cno}---for a set of tracked spoofed signals---can then be utilized to detect spoofing. Similarly, a sudden decrease in AGC gain settings along with a decrease in \gls{cno} measurements can indicate jamming. In this situation, the jammer injects an excessive amount of signal uncorrelated with the authentic GNSS signals causing their \gls{cno} to drop while the AGC decreases its gain due to the overall increase in power. 

To leverage these measurements as a spoofing countermeasure, the joint behavior of AGC and \gls{cno} is categorized into distinct regions that indicate: 
\begin{enumerate}
    \item nominal receiver operation
    \item low signal level due to signal obstruction
    \item GNSS denial due to jamming
    \item GNSS deception due to spoofing
\end{enumerate}
%
These regions are typically discriminated in countermeasure algorithms based on thresholds set for both the AGC value and \gls{cno} value.
\begin{figure}[!htbp]
    \centering
    \includegraphics[trim=9cm 5cm 10cm 5cm, clip,width=\linewidth]{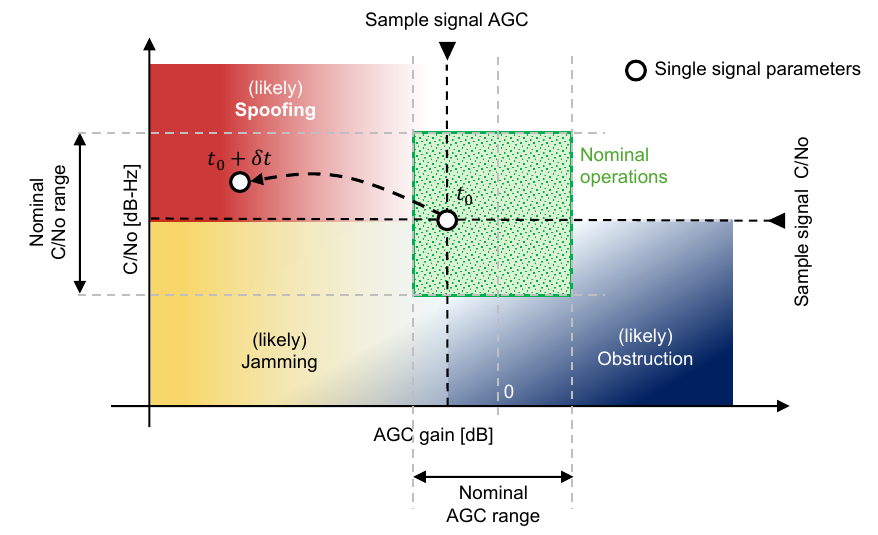}
    \caption{AGC vs. \gls{cno} signal parameters space along with conventional operating and critical regions.}
    \label{fig:cn0agcPlot1}
\end{figure}

Figure~\ref{fig:cn0agcPlot1} illustrates the aforementioned operating regions in 2-dimensional space. Among these, a nominal operational region is defined in green by the AGC dynamic range and a typical \gls{cno} dynamic range. The white dot represents a single set of signal parameters (AGC and \gls{cno}) associated to a specific time instant, $t_0$, for a given satellite signal, acknowledging that the AGC applies uniformly to all signals. We consider that a GNSS receiver handles multiple concurrent signals that share the same AGC value but exhibit different \gls{cno} levels, as commonly observed in real operational scenarios. 

Taking into account the temporal variations $t = t_0 + \delta t$, and under the hypothesis of a spoofing attack that affects both AGC and \gls{cno}, the dots in the set $\mathcal{O}^{S}$ may shift to the “spoofing region”. In contrast, the dots in the set $\mathcal{O}^{L}$ remain largely stable in \gls{cno}, despite responding to the overall decrease in AGC caused by the additional power received. The analysis of the behavior over time of each dot in the diagram can hence be used to infer a potential ongoing spoofing attack on a subset of signals (and minimally enable a countermeasure approach from category 1). This also provides a discrimination between legitimate and non-legitimate signals, thus enabling exclusion algorithms (countermeasure category 2). 

%
%
%
An early study implemented this method before the release of the Android API that exposes raw GNSS observables \cite{valles2015interference}. In this work the authors utilized root access to the phones and intercepted data pipes in the Linux kernel to obtain simultaneous access to the AGC and \gls{cno} measurements. They demonstrate the feasibility of using smartphone measurements of the AGC and \gls{cno} by presenting data collected from a Google Nexus 7 in the presence of both jamming interference and high power spoofed GNSS signals. These data showed that the AGC and \gls{cno} measurements shifted in the expected directions in each of these scenarios. After the introduction of the Android API in 2016 that exposes raw GNSS observables, many other researchers reproduced similar results without needing root access to their phones. Additionally, this metric has been incorporated into several research works using combined metrics for detection \cite{lee2019analysis,ceccato2018exploiting,valles2015interference,lo2019test,spens2022detecting,minetto2023detecting}.

The most prevalent drawback of this approach is the inter-device inconsistency of reported AGC values. Across the diverse Android ecosystem, the different implementations of GNSS chipsets and software produce disparate AGC outputs~\cite{morichi2026}. One study \cite{lee2021agc} highlighted several critical discrepancies: first, AGC values do not follow a consistent signed direction relative to gain changes (increasing on some devices while decreasing on others). Second, some devices report an absolute AGC gain while others report a relative AGC gain. Finally, the baseline absolute values are inconsistent across hardware. These compounding factors make it difficult to develop a universal, threshold-based algorithm. Moreover, the AGC mechanism operates on the aggregate received power rather than on individual signal components. As a consequence, AGC measurements do not enable discrimination among different GNSS signals or interference sources, which constitutes an additional limitation in its use as a mitigation method (though this can be obtained by exploiting \gls{cno} data).

\subsection{Temporal Correlation-based Methods}
\label{sec:temp_corr_sec}

A different but related approach using \gls{cno} data was developed by \cite{minetto2023detecting,rustamov2023improving} where the time correlations between different satellites' \gls{cno} estimates were used as an indicator of spoofing signals transmitted from a single antenna. This approach exploits the channel diversity difference between a single antenna spoofer and authentic live sky signals. In fact, a single antenna spoofer emits all its satellite RF signals from a single emitting source.

The received signal under nominal conditions (that is, receiving only legitimate satellite signals) was defined earlier in \eqref{eq:sig_mod} and its spoofed counterpart in \eqref{eq:sig_mod_spoofing}. In particular, the legitimate noiseless received signal $x(t;\bm{\theta})$  there can be interpreted as 
\begin{equation}
x(t;\bm{\theta}) = \sum_{i\in\mathcal{I}^L} x_i(t) * h_i(t) \;,
\label{eq:received_signal3}
\end{equation}
where it is made explicit that the received signal is a version of the transmitted signal $x_i(t)$ that is modified by the propagation channel, whose impulse response is different for each satellite and denoted by $h_i(t)$. This model represents the superposition of $M$ satellite signals, propagating through mostly independent channels. The $i$-th transmitted signal is defined as $x_i(t) \doteq x_i(t;\bm{\theta}_i^0)$, with $\bm{\theta}_i^0=(P_i^{\textrm{T}},0,0,\phi_i^0)^\top$, while the corresponding received component is $x_i(t;\bm{\theta}_i)=x_i(t)*h_i(t)$. 

In the presence of spoofing (or meaconing) via a single antenna, the combined spoofed received signal in \eqref{eq:sig_mod_spoofing} can be interpreted as
\begin{equation}
x^S(t;\bm{\theta}^S)
= \Bigg[
\sum_{j\in\mathcal{I}^S} x_j^S(t;\bm{\theta}_j^{S}) + \eta^S(t)
\Bigg] * h^{S}(t)
\label{eq:spoofed_signal3}
\end{equation}
where $x_j^S(t;\bm{\theta}_j^{S})$ denotes the forged signal for the $j$-th satellite in the spoofing list $\mathcal{I}^S$. The received signal is not a simple combination of those signals, but there are other two distinctive aspects to its reception: $h^{(S)}(t)$ denotes the common propagation channel between the spoofer and the target receiver, and $\eta^S(t)$ random noise possibly introduced by the attacker's front-end, as described also in Section~\ref{sec:rx_basics}. 
This formulation highlights two key aspects: $i)$ all spoofed signals share the same channel, inducing correlated distortions despite their individually assigned parameters $(P_i,\tau_i,f_{Di},\phi_i)^\top$; and $ii)$ the spoofer noise undergoes the same propagation channel. It is important to note that while the spoofer computes its own parameters in $(P_i,\tau_i,f_{Di},\phi_i)^\top$, the signal it emits is convolved with $h^{S}(t)$ which precludes the spoofer from controlling the exact parameters received by the victim receiver, $\bm{\theta}^{S}$.
Equations \eqref{eq:received_signal3} and \eqref{eq:spoofed_signal3} thus provide a unified representation of the received signal (before adding the receiver's thermal noise $\eta(t)$) under nominal and spoofed conditions. This results in the temporal correlations between the spoofed satellite signals being higher than the correlations in authentic satellite data. 

By exploiting Pearson correlation coefficients to identify pairwise correlation among time series, the analysis becomes independent of specific smartphone chipsets. Once short time-series of a few seconds are collected, a detection threshold can be established either using a large set of training datasets or by leveraging Neyman–Pearson detection theory, which determines the threshold based on a specified probability of false alarm. Pairwise correlation analysis also enables the identification of pairs of potentially spoofed signals, allowing their selective exclusion from the navigation solution (countermeasure category 2). 

\begin{figure*}[!htbp]
    \centering
    
    \begin{subfigure}[t]{0.48\linewidth}
        \centering
        \includegraphics[width=\linewidth]{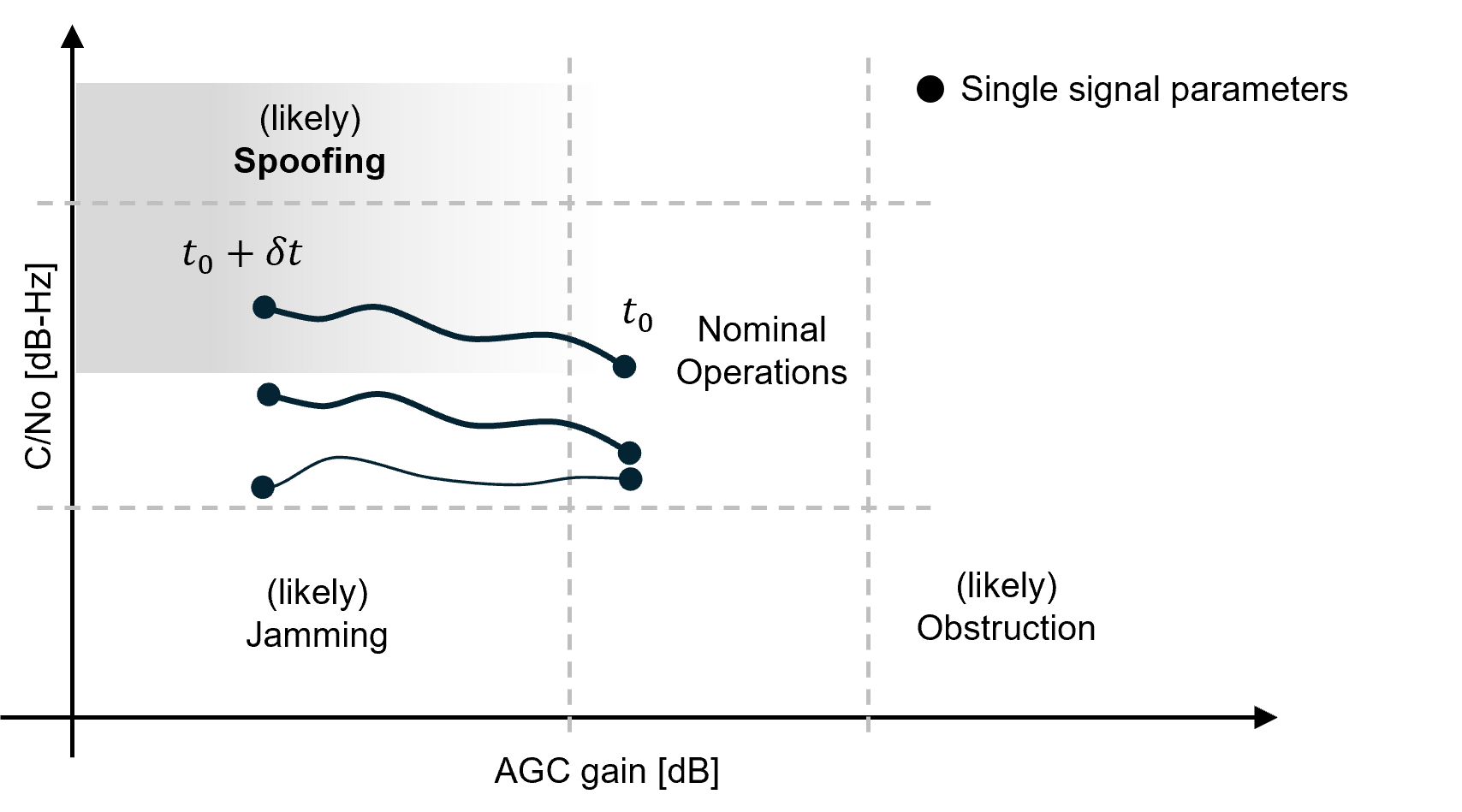}
        \caption{C/No vs AGC for multiple signals over time}
        \label{fig:sub1}
    \end{subfigure}
    \hfill
    \begin{subfigure}[t]{0.46\linewidth}
        \centering
        \includegraphics[width=\linewidth]{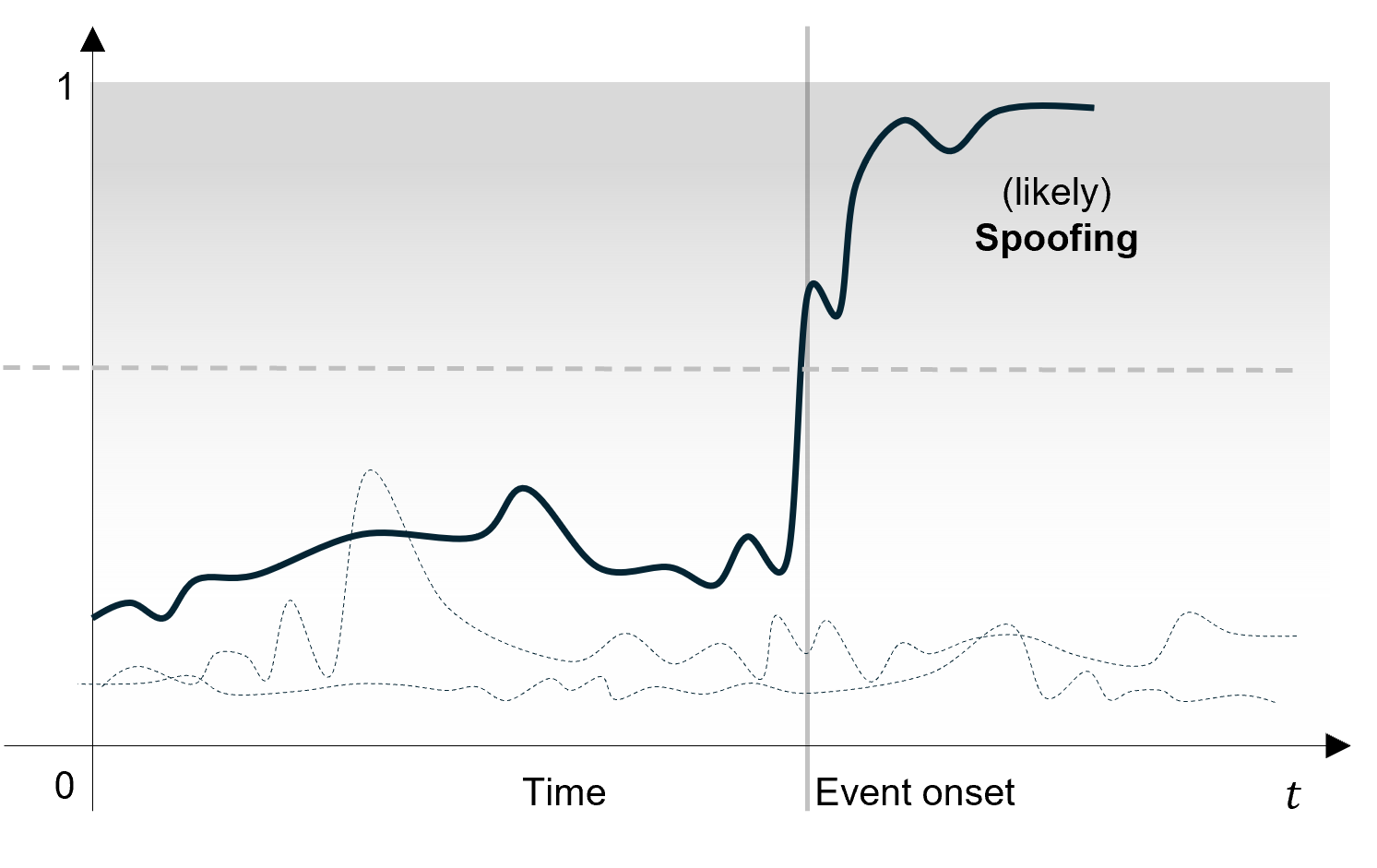}
        \caption{Correlation coefficient}
        \label{fig:sub2}
    \end{subfigure}
    
    \caption{Augmented signal parameters space with pairwise time series correlation dimension.}
    \label{fig:timeCorrelationextension}
\end{figure*}

This concept is depicted in Figure~\ref{fig:timeCorrelationextension}, and it  extends the paradigm of Figure~\ref{fig:cn0agcPlot1} by introducing a further dimension in the signal parameters space: a temporal correlation coefficient. This coefficient can be computed for any parameter set experiencing temporal correlations; here, without loss of generality, we illustrate the concept using \gls{cno} time series data. Under non-spoofed conditions, pairwise measurements from individual GNSS satellites at $t=t_0$ exist in their nominal measurement ranges. However, when affected by spoofing at $t=t_0+\delta t$, these points shift toward the ``spoofing-like'' region. By extending the signal parameter space to include temporal correlation, this shift is further validated; the time-series correlation relocates the pairwise points firmly within the spoofing region defined for this dimension in Figure~\ref{fig:sub2}.

The paradigm can be expanded to other set of GNSS observables and complementary features can be leveraged once direct effect can be demonstrated under spoofing. Multidimensional feature analysis can also be used to identify informative clustering effects in the observable domain.

This class of approaches is generally more robust than methods based solely on AGC and \gls{cno} measurements. Using channel diversity and exploiting Pearson correlation coefficients enables the use of metrics that have a more uniform behavior under spoofing, and the analysis becomes independent of specific smartphone chipsets. Once short time-series of a few seconds are collected, a detection threshold can be established either using a large set of training datasets or by applying the Neyman–Pearson detection framework, which determines the threshold based on a specified probability of false alarm. Furthermore, pairwise correlation analysis enables the identification of pairs of potentially spoofed signals, allowing their selective exclusion from the navigation solution.

\subsection{Inertial-based Methods}
\label{sec:inert_sec}
Inertial-based countermeasure methods utilize inertial sensors co-located with the GNSS receiver to detect spoofing. Inertial sensors measure a different modality (e.g. acceleration) from GNSS signals. Since this alternative measurement modality is unaffected by GNSS spoofing, measurements from inertial sensors become inconsistent with measurements from GNSS receivers under GNSS spoofing. This inconsistency can be exploited as a countermeasure technique.

Inertial-based detection methods are a natural focus area in the context of smartphones, since these devices contain inertial measurement units as a part of their sensor suite. The quality of inertial sensors on smartphones are limited due to cost, but the amount of drift in their measurements over short periods of time (i.e. a few seconds) is small enough that they are still useful in the development of counteremeasure techniques. Several researchers have pointed out the possible utility of using inertial sensors on smartphones as a potential spoofing detection method. In \cite{miralles2018android}, the authors investigate the feasibility of using inertial data by presenting smartphone inertial measurements alongside GNSS PVT measurements during a spoofing attack and during benign conditions. The authors point out that there is a discernible difference between the 2 scenarios in the data, though they do not present a detection algorithm using these data. Furthermore, they concede that this approach may be difficult on smartphones due to the limited quality of their inertial sensors. Similarly, \cite{ceccato2018exploiting} investigated the feasibility of using smartphone inertial sensors to detect spoofing by estimating the cumulative distribution functions (CDFs) of the sensor output in different dynamics environments. The authors did not directly apply these statistical characterizations to a spoofing detection algorithm, but commented that this approach could have utility in this context through rough cross checking.

In \cite{lee2019analysis}, the authors use carrier phase measurements from the raw GNSS measurement API to estimate parameters such as attitude rate, heading velocity, and acceleration that can be compared to independently-derived estimates of the same parameters from the inertial sensors onboard the phone. For a spoofing detection metric, the authors compared the mean values between inertial-derived estimates and GNSS-derived estimates over a chosen time period. The authors test 2 different smartphones in scenarios without a spoofer present. Under these benign conditions, they compare the differences between the GNSS-derived and inertial-derived estimates of the heading rate, acceleration, and attitude. A low false alarm rate was achieved in the testing scenario for each of these metrics; however, the study did not test the method against data with spoofing signals present. Furthermore, the authors note that in order to achieve a low false alarm rate, it is likely a different threshold will need to be derived per-device to accommodate the different inertial sensors embedded in different model smartphones.

To the best of our knowledge, \cite{zhou2024performance} provides the most comprehensive development of an inertial-based spoofing countermeasure for smartphones. In this study, the authors develop a detection method based on the innovation of a loosely-coupled fusion model. The authors fuse inertial measurements with GNSS measurements using a Kalman Filter. In this context, $\Xi \left(\cdot\right)$ in equation \eqref{state_estimate} from section \ref{sec:pvt} becomes the function defined by the Kalman Filter. For this particular study, the authors chose the following formulation for their state vector 
$\bm{x}_k=\begin{bmatrix}
\delta\bm{p},
\delta\bm{v},
\delta\bm{\phi},
\bm{b}_a,
\bm{b}_g
\end{bmatrix}^{\top}$
where $\delta\bm{p}$, $\delta\bm{v}$, and $\delta\bm{\phi}$ represent the error estimation of position, velocity, and user attitude respectively, and $\bm{b}_a$ and $\bm{b}_g$ represent the bias in accelerometer and gyroscope, respectively. The innovation vector is typically used as a detection metric, defined as the difference between the observations and the predicted measurements based on the predicted state vector $\hat{\bm{x}}_{k|k-1}$. 
%
Deviations between GNSS and inertial measurements manifest themselves in the innovation vector. The authors then distinguish the distributions of the innovation vector under two scenarios: one with spoofing and one without spoofing. The distributions are used in a binary hypothesis test with a threshold set by a desired false alarm rate. 

This detection method was applied to 3 different smartphones in multiple spoofing scenarios using a back-end machine to perform the processing. The researchers found this method effective in detecting spoofing signals; in the scenarios where the phones were subject to a strong spoofing signal, the three phones reliably reported innovation vector values above their set threshold indicating spoofing. The authors conceded that one challenge of this approach is keeping the false alarm rate low. This method operates under the assumption that in an environment without spoofing, there will be a high level of self-consistency between the inertial and GNSS measurements. The high level of consistency should then result in small values of the innovation vector. In reality, the innovation vector values also increased in scenarios with high multipath and weak GNSS signal strength. Furthermore, the innovation vector had slightly different distributions between the different phones due to the different sensors and processing embedded in the phones.

\subsection{Crowdsource Methods}
\label{sec:crowd_sec}
Crowdsourced methods refer to the concept of using the collective measurements of many co-located smartphones to improve interference detection and localization performance \cite{strizic2018crowdsourcing}. This approach leverages the advantage of having many independent sensor measurements of a particular geographic area to drive down false alarm rates and estimate area of effect. This is particularly useful for smartphones that tend to have a noisy, heterogeneous measurement space from their low-cost, diverse set of sensors. Multiple studies have proposed crowdsourcing methods that can be used to this end. In \cite{valles2015interference} the authors propose a multi-user approach that utilizes $C/N_0$ values from the phones to detect interference events. In this approach, if all users in a minimal geographic area experience a drop in signal power (as reported by the $C/N_0$), a jamming event is recorded. Conversely, if a minority of users in some area report a signal power drop while others do not, the event is reported as a signal blockage (some obstruction moves between the user(s)' phone and satellites). The authors tested both scenarios and present data showing the feasibility of using this approach in a networked setting. In the jamming scenario the authors were able to detect the presence of the jamming signal but also created multiple false alarms. In the signal blockage scenario. the detection algorithm successfully distinguished a blocked signal from a jamming event, but misclassified signal blockage when a majority of the devices simultaneously lost line-of-sight to the satellites (e.g. multiple users walking into a building).

A later study investigated a similar approach utilizing the raw GNSS measurement API to expand the number of methods a crowdsourced approach could utilize \cite{lo2019test}. In this study the authors explore three different metrics to detect spoofing: position/velocity, acceleration, and signal power. A combination of a spoofing field test and measurements taken in benign environments were used to assess the feasibility of each approach. The smartphones used in the spoofing field test did not report measurements from the raw GNSS measurement API, but the authors present position, acceleration, and $C/N_0$ data. For each of these measurements there are discernible differences between the spoofed and benign signal environments, but the authors do not present a unified detection metric or network-based algorithm.

Multiple raw measurement types available in smartphones can indeed act as proxies for spoofing signal power. In particular, AGC levels and $C/N_0$ measurements are effective indicators of anomalously strong signals, which are typically required for successful spoofing. Moreover, because the received interference power is inherently related to the distance from the spoofer---as dictated by propagation effects---crowdsourced interference power measurements can be used to support not just emitter detection but also localization.

This concept has already been demonstrated in the context of jamming attacks~\cite{olsson2022participatory,raichur2022machine,nardin2023jloc,nardin2023crowd,jaramillo2025jamloc,7439734}. The same general methodology can be ported to spoofing scenarios once the relationship between received spoofing power and measured observables is characterized. While the mapping between observables (e.g., AGC, $C/N_0$) and actual interference power differs between spoofing and jamming (see Fig.~\ref{fig:cn0agcPlot1}), the underlying localization algorithms remain applicable. Moreover, a practical challenge in crowdsourced environments is the heterogeneity of devices: noise figures, AGC implementations, and $C/N_0$ estimation methods vary across smartphones. For this reason, deriving an absolute and device-independent interference power metric is essential to place measurements from different users onto a common scale. Achieving such normalization generally requires calibration and cross-device harmonization of observables~\cite{bastide_automatic_2003,borio_jammer_2016}.

\begin{figure}[!htbp]
    \centering
    \begin{subfigure}[t]{\columnwidth}

    \includegraphics[trim=0cm 0cm 0cm 0cm, clip,width=\linewidth]{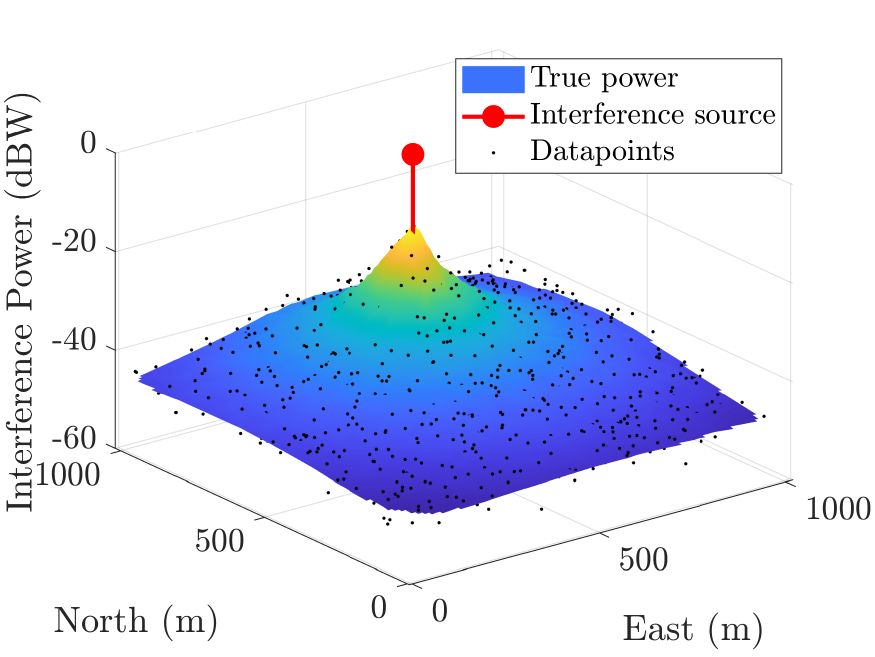}
    \caption{Soft information approach: Interference power field estimation.}
    \label{fig:softinfo}
    \end{subfigure}
\hfill
\begin{subfigure}[t]{\columnwidth}
    \includegraphics[trim=0cm 0cm 0cm 0cm, clip,width=\linewidth]{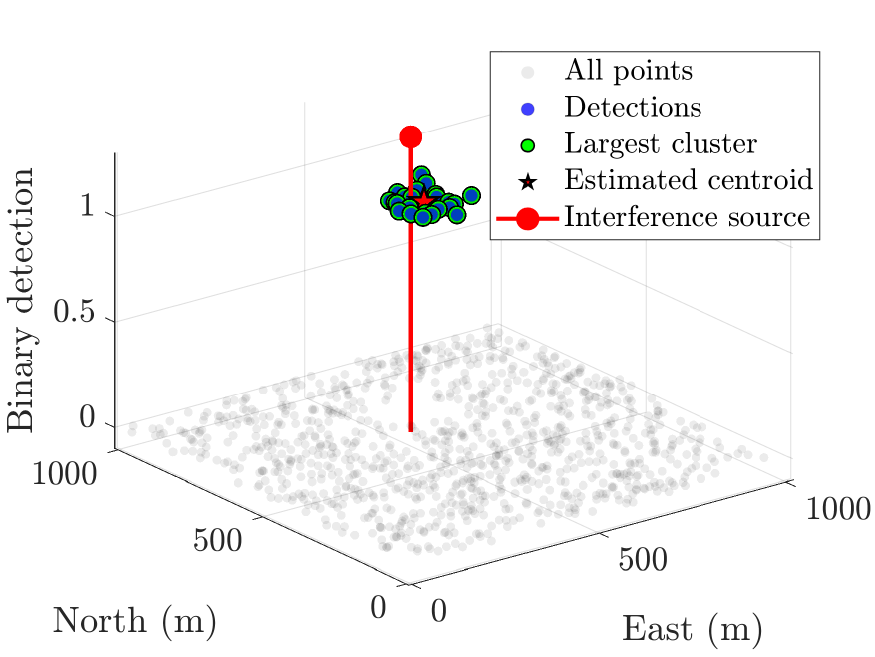}
    \caption{Hard information approach: Binary detection based on estimated interference power.}
    \label{fig:hardinfo}
\end{subfigure}
\caption{Crowdsourced interference detection and localization in open-field scenario (path loss propagation dominating).}
\label{fig:crowsourced_localization}
\end{figure}

Once interference power is estimated, it can be exploited in two ways. As soft information, continuous-valued interference power measurements can be directly modeled using propagation-based formulations \cite{gattis2025stress} and data-driven models to jointly estimate the interference source location and spatial power field~\cite{nardin2023jloc,nardin2023crowd}. This approach can also leverage distributed modeling methods, like federated learning, to preserve privacy of crowdsourced devices~\cite{jaramillo2025jamloc}. Alternatively, interference power can be converted into hard information through thresholding or classification, yielding binary detection data. In this case, localization can follow a simple two-step approach where a density-based clustering (e.g., DBSCAN~\cite{ester1996dbscan,schubert2017dbscanrevisited}) can be used to discard isolated detections, and a centroid or geometric mean computation of the dominant cluster is employed to estimate the interference source position. The latter approach is computationally lightweight and leaves room for more advanced probabilistic refinements. Examples of both strategies are illustrated in Fig.~\ref{fig:crowsourced_localization} for an ideal open-field scenario dominated by path-loss propagation, while more realistic urban environments have been analyzed for the jamming case in~\cite{nardin2023jloc,nardin2023crowd,nardin2023innovative,jaramillo2025jamloc}.

In \cite{rustamov2020gnss} the authors investigate the implications of a spoofing attack on a network of GNSS receivers (i.e., smartphones) exchanging raw measurement data for the purpose of improving positioning accuracy using differential GNSS (DGNSS). In this case, a spoofer that captures one device in the network has the opportunity to corrupt the positioning solution of other devices in the network that are ingesting the captured device's raw measurements. The authors explore this threat through a combination of RF testing and simulation. In their RF testing, they spoof several smartphones in an open-air scenario where the spoofed signals are contending with real satellite signals. They found that even when the phones did not report the spoofed PVT position, they still tracked some of the spoofed satellite signals. This is significant for this scenario under study because raw measurements are being exchanged between receivers, which can still be corrupted even if the receiver's position solution remains accurate after excluding these measurements from its position calculation. The simulated cases studied the effect of corrupted measurements from an aiding device on another device using DGNSS. The results showed observable discontinuities in the receiver's solution performing DGNSS, and recommended a threshold-based edge detection mechanism as a possible solution for identifying spoofed data in the networked measurements exchanged.
Finally, the authors in \cite{miralles2019gnss} present an approach for integrating smartphones into the GNSS Threat Reporting and Receiver Testing through International Knowledge Exchange, Experimentation and Exploitation (STRIKE3) Threat Monitoring and Reporting standard. STRIKE3 is a standard under the European Union attempting to formalize the process of threat reporting and monitoring within the GNSS space. The authors propose changes to the STRIKE3 standard that would support raw GNSS measurements from smartphone devices that could supplement the higher-end receivers that currently make up that network.

\subsection{Multi-Data Source Methods}
\label{sec:multi_data_sec}
A natural extension of the countermeasure techniques explored so far is combining many of these metrics into a single detection algorithm that assesses a more holistic set of data before making a decision. Comparison of the PVT solution obtained from GNSS data in the smartphone with other network-based solutions offers an attractive detection strategy that shows promise~\cite{formaggio2019gnss}. \cite{ceccato2018exploiting} proposed several possible approaches in their research to detect GNSS spoofing in smartphones, including position history, comparison of GNSS-derived PVT with network-derived PVT, and cross-checking with mobile sensors. However, in this study, the researchers did not propose a detection method integrating each of these approaches. Around the same time researchers out of the University of Colorado Boulder revealed they were developing an Android app called GNSSAlarm, which integrates several detection metrics into a single app to detect GNSS spoofing \cite{miralles2018android,miralles2020robust}. A few years later this app was presented in a more mature form, and used four different measurements: (1) comparing GNSS and network locations, (2) checking the Android mock location flag, (3) comparing GNSS and system times, and (4) monitoring the AGC and $C/N_0$ values \cite{spens2021application}. The researchers use a moderate position difference between GNSS and network positions as a threshold for indicating spoofing (200 meters) due to the limited accuracy of network-based positioning. The mock location flag offers a simple boolean measure to indicate if a user application is spoofing the GNSS position through non-RF means. A timing difference of 2 seconds was used between GNSS and network time to accommodate the update cadence of the GNSS measurements. The AGC and $C/N_0$ combined metrics investigated expected trends in the presence of jamming or spoofing, but the authors concede this approach likely requires parameter tuning specific to different phones. These detection metrics are combined in a single smartphone app that indicates the likelihood of spoofing based on the number of metrics indicating a discrepancy. In other studies the researchers explored additional metrics as candidates to add to their spoofing detection suite, including inertial data and pseudorange residuals \cite{miralles2018android}. However, it remains unclear whether and how these metrics are integrated into the current implementation of the GNSSAlarm app.

\subsection{Multi-Frequency Methods}
\label{alt_sig_sec}
An increasing number of smartphones process GNSS signals from multiple constellations, and even multiple frequency bands to improve positioning performance. Less sophisticated spoofers may emit in a single GNSS band (e.g. L1), and perhaps only spoof GPS C/A code. Spoofing detection algorithms can take advantage of this lack of sophistication by comparing GNSS measurements between different frequency bands, such as the L1 and L5 bands. In this case, the major deviations of the measurements received between the L1 and L5 bands would be a clear indicator of spoofing. Furthermore, the spoofing signal could be excluded from the positioning solution and the L5 signal could provide the smartphone's true location. This would offer mitigation to the spoofing signal and not just detection (countermeasure category 2). \cite{mcburney2024empirical} explored this topic by looking at collected data from real-world spoofing events that took place and \cite{blum2022smartphone} did earlier with smartphone data. The researchers found that even many GNSS chips that support both L1 and L5 frequencies are spoofed from an L1-only spoof. They posit this is due to the legacy algorithm many chips employ, which rely on L1 signals to acquire and track L5 signals. The researchers argue that a chip that acquires and tracks L5 signals independently from L1 signals will be less susceptible to spoofing. Indeed, the researchers demonstrate this by showing their GNSS chip design using this methodology remains unspoofed in these real world events. While this method does not necessarily protect against spoofers emitting in multiple frequency bands, it does increase the minimum level of sophistication a spoofer must employ to successfully deceive its target.

\section{Critical Comparison and Technological Prospects}
\label{sec:critical_comparison}
\subsection{Countermeasure Comparison and Assessment}
Table~\ref{tab:concl_table} provides a comparative assessment of the smartphone-specific countermeasure families reviewed in this section along three dimensions: the anti-spoofing capability they achieve, their principal limitations, and their technological maturity. This comparison reveals that no single method dominates across all three dimensions. Instead, the available approaches occupy different positions in a trade space involving deployability, device independence, attack coverage, and attainable protection.

\begin{table*}[t]
\caption{Summary of existing smartphone-specific countermeasure methods, limitations, and maturity}
\label{tab:concl_table}
\centering
\renewcommand{\arraystretch}{1.3}
\setlength{\tabcolsep}{4pt}

\begin{tabularx}{\textwidth}
{c c c c c c}
\toprule
\textbf{Countermeasure Method} & 
\textbf{Section Number} & 
\textbf{\makecell{Countermeasure \\ Category}} & 
\textbf{Limitation}& 
\textbf{Maturity} & 
\textbf{References}\\
\midrule

Combined AGC \& $C/N_0$ Metric 
& \ref{sec:agc_cn0_sec} 
& \blocks{1,0,0}
& \blocks{1,0,0,0,0} 
& \gauge{2} 
& \makecell{\cite{lee2019analysis}, \cite{ceccato2018exploiting}, \cite{valles2015interference},\\ \cite{lo2019test}, \cite{spens2022detecting}, \cite{minetto2023detecting}, \cite{lee2021agc}}\\

\midrule
Temporal Correlation-based Methods 
& \ref{sec:temp_corr_sec} 
& \blocks{1,1,0}
& \blocks{1,0,0,0,0} 
& \gauge{1} 
& \cite{minetto2023detecting,rustamov2023improving} \\

\midrule
Combined GNSS/Inertial Methods 
& \ref{sec:inert_sec} 
& \blocks{1,0,0}
& \blocks{1,1,0,0,0} 
& \gauge{1} 
& \cite{miralles2018android}, \cite{ceccato2018exploiting}, \cite{lee2019analysis}, \cite{zhou2024performance}\\

\midrule
Crowdsource Measurements 
& \ref{sec:crowd_sec} 
& \blocks{1,0,0}
& \blocks{1,0,1,0,0} 
& \gauge{0} 
& \cite{valles2015interference}, \cite{lo2019test}, \cite{rustamov2020gnss}, \cite{miralles2019gnss}\\

\midrule
Multi-Data Source 
& \ref{sec:multi_data_sec} 
& \blocks{1,0,0}
& \blocks{1,0,0,0,0} 
& \gauge{2} 
& \cite{ceccato2018exploiting}, \cite{miralles2018android}, \cite{miralles2020robust}, \cite{spens2021application}\\

\midrule
Multi-Frequency Methods 
& \ref{alt_sig_sec} 
& \blocks{1,1,0}
& \blocks{0,0,0,1,1} 
& \gauge{2} 
& \cite{mcburney2024empirical}\\

\bottomrule

\multicolumn{6}{@{}p{3in}}{\textbf{Countermeasure Category Key}}\\
\multicolumn{6}{@{}p{7in}}{\blocks{1,0,0} ASC1 \qquad \blocks{1,1,0} ASC2 \qquad \blocks{1,1,1} ASC3 } \\
\hline
\multicolumn{6}{@{}p{1.5in}}{\textbf{Limitation Key:}}\\
\multicolumn{6}{@{}p{7in}}{\blocks{1,0,0,0,0}  Difficult to generalize across a diverse set of devices}\\
\multicolumn{6}{@{}p{7in}}{\blocks{0,1,0,0,0}  Implementation may have a high false alarm rate in more complicated RF environments (e.g., high multipath)}\\
\multicolumn{6}{@{}p{7in}}{\blocks{0,0,1,0,0} Is most effective when widely adopted by a critical mass of devices}\\
\multicolumn{6}{@{}p{7in}}{\blocks{0,0,0,1,0} Requires device able to process multiple signals in different frequency bands}\\
\multicolumn{6}{@{}p{7in}}{\blocks{0,0,0,0,1} Most effective for spoofers with limited sophistication (i.e., single constellation and/or frequency band)}\\
\hline
\multicolumn{6}{@{}p{1.5in}}{\textbf{Maturity Key:}}\\
\multicolumn{6}{@{}p{7in}}{\gauge{0} Methodology for implementation proposed, initial simulations conducted to assess feasibility}\\
\multicolumn{6}{@{}p{7in}}{\gauge{1} Proof of concept implemented and initial RF experiments conducted to demonstrate feasibility}\\
\multicolumn{6}{@{}p{7in}}{\gauge{2} Software and/or hardware available to purchase and demonstrated in realistic scenarios}\\

\end{tabularx}
\end{table*}

A first important distinction concerns the level of protection provided. Most of the reviewed methods belong to ASC1: they can indicate that spoofing is likely present but cannot determine which measurements are compromised. Only temporal-correlation and multi-frequency methods reach ASC2 under the assumptions considered, because they can potentially identify correlated signals or isolate measurements received on an affected frequency. None of the reviewed method families currently reaches ASC3, which would require excluding spoofed observations and restoring a trusted PVT solution. This gap is particularly consequential for smartphones. An application may detect or even identify spoofed measurements without having sufficient control over the chipset or operating-system location service to alter the native PVT solution. Therefore, successful application-level detection does not necessarily translate into platform-level protection.

Combined AGC and $C/N_0$ monitoring is among the most mature and readily deployable approaches because it uses measurements available through Android APIs and requires limited computational resources. However, device-dependent AGC conventions, baseline values, and $C/N_0$ estimation procedures make universal thresholds difficult to establish. Moreover, similar power anomalies may be caused by jamming, blockage, or propagation effects. This method therefore provides a useful first warning layer but is insufficient as a stand-alone spoofing indicator.

Temporal-correlation methods exploit the common temporal behavior typically introduced when counterfeit satellite signals originate from a single antenna. By relying on relative rather than absolute measurements, they can reduce sensitivity to device-specific baseline values and identify groups of potentially spoofed signals, thereby reaching ASC2. Their effectiveness nevertheless depends on assumptions about the spatial structure of the attack and may decrease against more sophisticated transmitters. Thresholds and observation windows also require validation across devices and environments, while the available experimental evidence remains less mature than for AGC and $C/N_0$ monitoring.

GNSS/inertial consistency methods compare two physically distinct sensing modalities and can reveal inconsistencies caused by deceptive position or velocity trajectories. Their effectiveness is nevertheless limited by consumer-grade sensor biases and drift, arbitrary device motion and orientation, and benign GNSS errors that may produce similar inconsistencies. Furthermore, at the time of writing, methods based on GNSS/inertial consistency remain at the proof-of-concept stage, with validation limited to experimental results. Multi-data-source methods combine signal-power, time, position, network, and sensor-consistency indicators to compensate for these individual weaknesses. Although comparatively mature, their performance still depends on the availability, calibration, and statistical dependence of the constituent indicators; robust fusion therefore requires accounting for their reliability under both attacks and benign anomalies.

Crowdsourced methods use spatially distributed measurements for collective interference detection, characterization, and localization. This spatial diversity can distinguish isolated device anomalies from geographically coherent events, making crowdsourcing promising for large-scale monitoring. However, Table~\ref{tab:concl_table} identifies it as the least mature family. Its deployment requires a sufficient density of contributing devices, reliable communications, measurement harmonization, and appropriate mechanisms addressing latency, privacy and device trustworthiness. It should therefore be regarded as a complementary infrastructure-level capability rather than a self-contained defense for an individual smartphone.

Multi-frequency processing provides a comparatively mature single-device approach to identifying affected measurements when an attack does not cover all frequencies processed by the receiver. Cross-frequency inconsistencies can reveal such attacks. However, its effectiveness depends on hardware support, independent processing of the different bands, and access to the required measurements. A coordinated multi-frequency and multi-constellation attack can remove this diversity; the method therefore raises the technical cost of spoofing but does not provide general immunity.

The maturity levels in Table~\ref{tab:concl_table} should thus be interpreted as indicators of implementation and experimental validation, rather than as measures of complete security. 
The most credible technological path is consequently a layered architecture combining complementary and, as far as possible, independent evidence: power and temporal features for rapid signal-domain warning, inertial and multi-data source information for PVT consistency, multi-frequency measurements for signal identification, and crowdsourced observations for spatial confirmation and source localization.

Beyond comparing the individual methods, Table~\ref{tab:concl_table} reveals a structural limitation of current smartphone countermeasures: none reaches ASC3 according to the definitions adopted in this survey. The surveyed approaches rely primarily on observations at later processing stages and other information available to applications, which can support the detection and, in some cases, identification of spoofed signals, but generally provide neither the information nor the control required to remove compromised measurements from the native receiver solution. Table~\ref{tab:methods}, which summarizes general spoofing countermeasures, supports this interpretation: methods operating on more information-rich data at earlier stages of the receiver processing chain can implement a broader range of anti-spoofing actions, whereas post-correlation methods do not attain ASC3. The most direct path toward mitigation is therefore the native integration of anti-spoofing mechanisms by chipset manufacturers, where access to interference-characterization evidence and control over receiver processing are greatest (see Fig.~\ref{fig:rx_stages}). The operating system may provide a further protection layer by accessing intermediate receiver information and combining it with inertial, network-based, and other positioning sources, without necessarily exposing those measurements to third-party applications. Selected information could also be made available through controlled APIs for specialized applications, although this may increase the attack surface and transfer part of the responsibility for validating GNSS outputs to application developers. The future effectiveness of smartphone countermeasures therefore depends not only on improved detection algorithms, but also on receiver architectures that provide sufficient observability and controlled intervention capabilities while avoiding the creation of additional security vulnerabilities.

\subsection{Open-Access Smartphone Spoofing GNSS Dataset}\label{sec:dataset}
Availability of datasets is of crucial importance in developing, testing, and benchmarking countermeasures (for instance, the popular TEXBAT dataset \cite{humphreys2012texas}). However, no available smartphone data was released to date to address the needs and challenges discussed in this article, for which we encourage researchers to refer to our open-source dataset of smartphone GNSS raw measurements as a means to continue development of novel countermeasure techniques~\cite{argo2026dataset}. This first-of-its-kind dataset includes smartphone data recorded from over-the-air experiments in both benign and spoofed GNSS environments, allowing researchers to examine device behavior across distinct operational environments. By incorporating numerous smartphone models and chipsets, the dataset enables direct testing of developed countermeasure techniques against one of the most challenging aspects of the smartphone platform space: hardware and firmware heterogeneity. \cite{argo2025multi} details the dataset itself and the collection mechanisms utilized to obtain the data.

\section{Conclusions and Outlook}
In this paper, we conducted a survey of existing studies exploring the susceptibility of smartphones to GNSS spoofing. Many research groups across numerous independent tests have found significant susceptibility of smartphones to GNSS spoofing, but due to the rich albeit limited quality sensor suite on smartphones there is ample opportunity to utilize existing data exposed on these devices to harden them to GNSS spoofing. We presented our taxonomy categorizing the effects of spoofing attacks on GNSS receivers, and introduced a framework for assessing countermeasure techniques. We then examined different spoofing countermeasure techniques that researchers have developed specifically for smartphones. Table \ref{tab:concl_table} summarizes the detection techniques explored in this paper and discusses the limitations and the level of maturity of each approach.

\usetikzlibrary{positioning}
\newcommand{\greenCircle}{\tikz\node[circle, draw=green!60, fill=green!60, very thick, minimum size=2.5mm,yshift=-0.5mm]{};}
\newcommand{\yellowCircle}{\tikz\node[circle, draw=yellow!80, fill=yellow!80, very thick, minimum size=2.5mm]{};}
\newcommand{\redCircle}{\tikz\node[circle, draw=red!60, fill=red!60, very thick, minimum size=2.5mm]{};}

The most common challenge found in GNSS spoof detection across different methodologies is accommodating the diverse, heterogeneous sensors available across different models of phones. Statistical characterization of sensor values is challenging across different physical sensors and/or firmware pairings with these sensors. This makes countermeasure methods based on Likelihood Ratio Tests difficult to generalize. This is particularly impactful for the signal power-based and inertial-based countermeasure methods. Recent works (e.g. \cite{minetto2023detecting}, \cite{argo2025multi}) have started exploring different statistical metrics based on power correlations and broader sets of available observables rather than characterization of absolute observable values (e.g. power distributions), which may prove to be more robust across a wider set of devices. Alternate detection methods do not suffer from this particular challenge, such as comparing network time with GNSS time or by utilizing GNSS signals in alternate bands. These methods make different assumptions that can be overcome with spoofer sophistication (e.g. aligning spoofed time with current time, and including multiple GNSS signals in the spoofed signal, respectively). Each countermeasure method has shown promise at different levels of maturity and often by making different sets of assumptions about the scenario or sensor set. Future evaluations should compare these combinations across heterogeneous devices, realistic benign environments, and adaptive spoofing attacks using common metrics such as detection and false-alarm statistics, delay, residual PVT error, computational cost, and cross-device portability. Given the wide array of approaches available and their demonstrated feasibilities so far, there are rich prospects for continued development and advancement of effective countermeasure techniques in the smartphone space.



\balance
\bibliographystyle{ieeetr}
\bibliography{bibliography_file,closas-ref}





\end{document}